\documentclass[onecolumn,draftclsnofoot,12pt]{IEEEtran}

\usepackage{amsfonts}
\usepackage{times}
\usepackage{graphicx}
\usepackage{latexsym}
\usepackage{dsfont}
\usepackage{amssymb}
\usepackage{amsmath}
\usepackage{cite}
\usepackage{verbatim}
\usepackage{subfigure}
\newcommand{\figref}[1]{{Fig.}~\ref{#1}}

\def\bb0{{\mathbb{0}}}

\def\ba{{\mathbf{a}}}
\def\bb{{\mathbf{b}}}
\def\bc{{\mathbf{c}}}
\def\bd{{\mathbf{d}}}

\def\bf{{\mathbf{f}}}
\def\bg{{\mathbf{g}}}
\def\bh{{\mathbf{h}}}

\def\bn{{\mathbf{n}}}

\def\br{{\mathbf{r}}}

\def\bx{{\mathbf{x}}}
\def\by{{\mathbf{y}}}
\def\bz{{\mathbf{z}}}
\def\b0{{\mathbf{0}}}

\def\bA{{\mathbf{A}}}
\def\bB{{\mathbf{B}}}
\def\bC{{\mathbf{C}}}

\def\bF{{\mathbf{F}}}
\def\bG{{\mathbf{G}}}
\def\bH{{\mathbf{H}}}
\def\bI{{\mathbf{I}}}

\def\bN{{\mathbf{N}}}

\def\bQ{{\mathbf{Q}}}
\def\bR{{\mathbf{R}}}

\def\bU{{\mathbf{U}}}
\def\bV{{\mathbf{V}}}
\def\bW{{\mathbf{W}}}
\def\bX{{\mathbf{X}}}
\def\bY{{\mathbf{Y}}}
\def\bZ{{\mathbf{Z}}}

\def\cA{\mathcal{A}}

\def\cC{\mathcal{C}}

\def\cH{\mathcal{H}}
\def\cI{\mathcal{I}}

\def\cN{\mathcal{N}}

\def\cX{\mathcal{X}}
\def\cY{\mathcal{Y}}
\def\cZ{\mathcal{Z}}

\def\sf0{{\mathsf{0}}}

\def\Nt{{N_t}}

\allowdisplaybreaks

\usepackage{latexsym}
\usepackage{graphicx}
\usepackage{amsmath}
\usepackage{amssymb}
\usepackage{amsfonts}
\usepackage{epsfig}
\usepackage{cite}   
\usepackage{calc}
\usepackage{color}
\usepackage{theorem}
\usepackage{url}
\usepackage{subfigure}
\usepackage{cases}
\usepackage{verbatim}
\usepackage{array}
\usepackage{dcolumn}
\usepackage{multirow}
\usepackage{tabularx}
\usepackage{setspace}
\usepackage{algorithm}
\usepackage{algcompatible}
\usepackage{upgreek}
\usepackage{stmaryrd}
\usepackage{multicol}

\newtheorem{defn}{Definition}

\newcommand{\sref}[1]{{Section}~\ref{#1}}

\newcommand{\bpf}{\begin{proof}}
\newcommand{\epf}{\end{proof}}
\newcommand{\bdefn}{\begin{defn}}
\newcommand{\edefn}{\end{defn}}

\usepackage{color}

\newcommand{\pinv}[1]{\ensuremath{#1^{\dagger}}} 	

\newenvironment{proof}{ \textbf{Proof:} }{ \hfill $\Box$}

\def\bW0{{\mathbb{0}}}

\def\ba{{\mathbf{a}}}
\def\bb{{\mathbf{b}}}
\def\bc{{\mathbf{c}}}
\def\bd{{\mathbf{d}}}
\def\be{{\mathbf{e}}}
\def\bf{{\mathbf{f}}}
\def\bg{{\mathbf{g}}}
\def\bh{{\mathbf{h}}}

\def\bn{{\mathbf{n}}}

\def\br{{\mathbf{r}}}

\def\bx{{\mathbf{x}}}
\def\by{{\mathbf{y}}}
\def\bz{{\mathbf{z}}}
\def\b0{{\mathbf{0}}}

\def\bA{{\mathbf{A}}}
\def\bB{{\mathbf{B}}}
\def\bC{{\mathbf{C}}}

\def\bF{{\mathbf{F}}}
\def\bG{{\mathbf{G}}}
\def\bH{{\mathbf{H}}}
\def\bI{{\mathbf{I}}}

\def\bN{{\mathbf{N}}}

\def\bQ{{\mathbf{Q}}}
\def\bR{{\mathbf{R}}}

\def\bU{{\mathbf{U}}}
\def\bV{{\mathbf{V}}}
\def\bW{{\mathbf{W}}}
\def\bX{{\mathbf{X}}}
\def\bY{{\mathbf{Y}}}
\def\bZ{{\mathbf{Z}}}

\def\bWC{{\mathbb{C}}}

\def\cA{\mathcal{A}}

\def\cC{\mathcal{C}}

\def\cH{\mathcal{H}}
\def\cI{\mathcal{I}}

\def\cN{\mathcal{N}}

\def\cX{\mathcal{X}}
\def\cY{\mathcal{Y}}
\def\cZ{\mathcal{Z}}

\def\sf0{{\mathsf{0}}}

\def\Lch{{L_\mathrm{ch}}}
\def\Kf{{K_\mathrm{sbcr}}}
\def\Tt{{T_\mathrm{frm}}}
\def\Nt{{N_\mathrm{ant}}}
\def\NtS{{N^2_\mathrm{ant}}}
\def\Nrf{{M_\mathrm{RF}}}
\def\Dg{{N_\mathrm{grid}}}

\def \Tr {\mathrm{Tr}}

\def \mB {{\mathring{\bB}}}
\def \mC {{\mathring{\bC}}}
\def \mG {{\mathring{\bG}}}

\begin{document}
\title {Spatial Channel Covariance Estimation  
for Hybrid Architectures Based on Tensor Decompositions
}
\author{
Sungwoo~Park, Anum Ali, Nuria Gonz{\'{a}}lez-Prelcic,  
and Robert~W. Heath Jr.
\thanks{S. Park, A. Ali, N.  Gonz{\'{a}}lez-Prelcic, and R. W. Heath Jr. are with the Wireless Networking and Communication Group (WNCG), Department of Electrical and Computer Engineering, The University of Texas at Austin, TX, 78701 USA. (e-mail: \{swpark96,anumali,ngprelcic,rheath\}@utexas.edu).  }
\thanks{This work is supported in part by the National Science Foundation under Grant No. 1514275, and by a gift from Huawei Technologies.}
}
\maketitle

\begin{abstract} 
Spatial channel covariance information can replace full instantaneous channel state information for the analog precoder design in hybrid analog/digital architectures.
Obtaining spatial channel covariance estimation, however, is challenging in the hybrid structure due to the use of  fewer radio frequency (RF) chains than the number of antennas. 
In this paper, we propose a spatial channel covariance estimation method based on higher-order tensor decomposition for spatially sparse time-varying frequency-selective channels.
The proposed method leverages the fact that the channel can be represented as a low-rank higher-order tensor. We also derive the Cram\'er-Rao lower bound on the estimation accuracy of the proposed method.  
Numerical results and theoretical analysis show that the proposed tensor-based approach achieves higher estimation accuracy  in comparison with prior  compressive-sensing-based approaches or conventional angle-of-arrival estimation approaches. 
Simulation results  reveal that the proposed approach becomes more beneficial at low  signal-to-noise (SNR) region.  
 \end{abstract}

\section{Introduction}\label{sec:intro}

Hybrid analog/digital precoding uses a smaller number of RF chains to reduce the number of power-consuming devices like  analog-to-digital converters (ADCs) or digital-to-analog converters (DACs).
Consequently, the hybrid approach can reduce power consumption and implementation complexity in millimeter wave multiple-input-multiple-output (MIMO) systems \cite{ElAyach2014,Roh2014,HeathJr2015,Sohrabi2016} and massive MIMO systems \cite{Zhang2005a,VenkateswaranVeen2010::1,Adhikary2013}. 
The rate loss incurred by the hybrid architecture is insignificant for spatially sparse channels such as in millimeter wave systems or in suburban/rural areas in sub-6 GHz systems \cite{ElAyach2014,Roh2014,HeathJr2015,Sohrabi2016,Adhikary2013}. 

A main challenge in the hybrid architecture is to configure the analog and digital precoding stages. 
Many previous methods accomplish this task based on full CSI \cite{ElAyach2014, Sohrabi2016, Coma2018jstsp}.
These approaches require frequent estimation of the channel, obtained for example via sparse recovery techniques. 
An alternative is to use only long-term statistical knowledge such as that contained in spatial channel covariance matrices, for the analog precoder design  \cite{Park2017twc,Park2017tsp,Adhikary2013,Rial2015camsap}. 
Once the analog precoder is determined based on the spatial channel covariance, the digital precoder, of a much smaller dimension is designed by using instantaneous full CSI of the low-dimensional effective channel, i.e., the propagation channel combined with the analog precoder. 
While the accurate estimation of the full CSI is difficult for time-varying frequency-selective channels, the long-term statistical CSI can be efficiently estimated.
It was shown in \cite{Park2017twc,Park2017tsp,Adhikary2013,Rial2015camsap} that the hybrid precoding methods based on spatial channel covariance achieve spectral efficiency close to that of the hybrid precoding obtained from  full CSI when the channels are spatially sparse.

Although the use of the spatial channel covariance matrix helps the hybrid precoding design to be simpler and more  practical, the hybrid architecture makes it difficult to estimate the covariance matrix. Since there is no digital access to the outputs of every antenna, and only the signals combined in an analog way are available at baseband, it is difficult  to estimate the spatial covariance of the high dimension  channel. 
Different approaches have been suggested to solve the spatial channel covariance estimation problem under such an environment. 
In \cite{VenkateswaranVeen2010::1}, a least-squares-based covariance estimation method was proposed by using time-varying analog beamforming matrices.  Since the method does not exploit the sparse channel property, it is not an efficient method for  sparse channels, which are of our interest in this paper. 
The sparse ruler array in \cite{RomeroLeus2013::11,ShakeriArianadaLeus2012::2, Ariananda2012Compressive} and the coprime sensor array \cite{VaidyanathanPal2011::3} can omit measurements on some antenna elements by leveraging the fact that correlations between antenna elements are wide-sense stationary in spatial domain. 
Although these so-called compressive covariance sensing (CCS) methods  can reduce the number of RF chains,
the methods have a limitation on the configuration of the number of RF chains and antennas. 

The CCS methods were initially developed by using only a subset of antennas. It is, however,  possible to extend the work to general hybrid architecture where the analog part is composed of phase shifters and thus is represented as a dense matrix  \cite{Ariananda2012Compressive}. 
In this dense sensing matrix case, the CCS methods become closely related to typical compressive sensing (CS) techniques. 
 For example, in  \cite{Rial2015camsap}, the spatial channel estimation method was developed by adopting the time-varying analog combiners used in  \cite{VenkateswaranVeen2010::1}, which are dense matrices. Instead of the least-squares method, one of the well-known conventional vector-type CS techniques,  orthogonal matching pursuit (OMP), 
was adopted to exploit the sparse channel property. 
It is worthwhile to note that conventional vector-type CS techniques were  typically developed for channel estimation \cite{Alkhateeb2014JSTSP,PengCL2015,LeeTC2016} but can be extended to spatial channel covariance estimation as well. 
The vector-type CS techniques for covariance estimation, however, are outperformed by matrix-type CS techniques developed for so-called multiple measurement vector (MMV) problems  \cite{CotterRaoEnganDelgado2005::6, ChenTSP2006,Determe2016} that enable the joint spare recovery.
An advanced spatial channel covariance estimation method based on the MMV approach was proposed in
 \cite{Park2017twc_Cov} by applying time-varying sensing matrices and exploiting the Hermitian property of the covariance matrix. The performance of the CS-based methods, however, is acceptable only in the moderate or high SNR region. 

Besides the CS-based work, conventional AoA estimation methods such as the multiple signal classification (MUSIC) \cite{Schmidt1986} and the estimation of signal parameters via rotational invariance technique (ESPRIT) \cite{Roy1989} can also be applied to the spatial channel covariance estimation problem via some modification.
In the conventional AoA estimation methods for fully-digital architectures, the spatial channel covariance is directly calculated from the received signal vectors,  and the AoA is estimated from the obtained covariance matrix. 
A covariance estimation based on AoA estimation for the hybrid architecture requires the estimation process in the opposite direction. First, the covariance matrix of the low-dimensional baseband received signal is calculated. Second, the AoAs are estimated from the covariance of the baseband received signal vectors. Finally, the covariance of the actual  channel is reconstructed from the estimated AoAs. 
This basic approach has been adopted for  different scenarios with some variations  \cite{ChuangWuLie2015,Guo2017}.
This approach, however, has a weak point:  the estimation accuracy rapidly decreases as the number of channel paths increases  toward the number of RF chains.
In addition, the methods based on this approach do not work properly when the number of channel paths exceeds that of RF chains. 

In this paper, we propose a spatial channel covariance estimation method for the hybrid architecture over uplink time-varying frequency-selective channels. 
We consider  a time division duplex (TDD) system where the estimated covariance over uplink channels can be used for the downlink precoding design at a base station (BS). 
We represent the channel and the received baseband signal as higher-order tensors. Considering spatially sparse channels,  we use the fact that these higher-order tensors have a low tensor rank and their canonical polyadic decomposition (CPD) forms are unique up to a common permutation and scaling of columns under some mild conditions \cite{Kolda2009,Cichocki2015,Sidiropoulos2017}. 
We also analyze the theoretical performance  by adopting the performance metric in \cite{Park2017twc_Cov,DBLP:journals/corr/HaghighatshoarC15::9} that is associated with the dominant eigenvalues and their eigenspaces of the spatial channel covariance matrix. 
We will call this performance metric the relative precoding efficiency (RPE) in this paper.
After showing that the RPE is closely related to the mean squared error (MSE) of the AoA estimation, we derive Cram\'er-Rao lower bound (CRLB) for the AoA estimation and its associated bound for the performance metric. 
Using numerical results, we first show that the performance of the proposed method approaches the performance bound as SNR increases. 
We also show that the lower bound of the tensor-based method is lower than that of the MUSIC-based method, which provides insight into the benefits of using tensor-based methods. Simulations show that the proposed tensor-based method outperforms CS-based methods as well as MUSIC-based methods, and its gain becomes more significant in the low SNR regime. 

The rest of the paper is organized as follows. \sref{sec:preliminaires} briefly introduces the basics of high-order tensor algebra. \sref{sec:sys_model} provides a system and channel model by using tensor representations. 
\sref{sec:Cov_est_tensor} describes the proposed spatial channel covariance method. 
The performance metric is analyzed in \sref{sec:Performance_metric}, and the CRLB is derived in \sref{sec:CRLB}. 
In \sref{sec:Comparison_others}, the proposed tensor-based work is compared with prior work based on CS or MUSIC. 
Simulation results are presented in \sref{sec:simulation_results}, and conclusions are drawn in \sref{sec:conclusions}.
This paper is the journal version of  \cite{Park2018globalsip} with theoretical analysis added.  

{\it{Notation:}} We use the following notation throughout this paper: $a$ is a scalar, $\ba$ is a vector, $\bA$ is a matrix, and $\mathcal{A}$ is a tensor.
$\bA^{\mathsf{T}}$, $\bA^{\mathsf{C}}$,  $\bA^*$, and $\pinv{\bA}$ are  transpose, conjugate,  conjugate transpose, and Moore-Penrose pseudoinverse. 
$[\bA]_{i,:}$ and $[\bA]_{:,j}$ are the $i$-th row and the $j$-th column of the matrix $\bA$.
$\bA \otimes \bB$, $\bA \circledcirc \bB$, and  $\bA \odot \bB$ denote the Kronecker product, the Hadamard product, and  the column-wise Khatri-Rao product.    $\ba \circ \bb$ denotes the outer product, which is also known as the tensor product. $\mathrm{Re}(\cA)$ and $\mathrm{Im}(\cA)$ denote the real part and the imaginary part of $\cA$.  $\mathrm{diag}(\bA)$ is a column vector whose elements are composed of the diagonal elements of $\bA$. 

\section{Preliminaries:  overview of  tensor algebra and canonical polyadic decomposition}\label{sec:preliminaires}

In this section, we review the basics of tensor algebra that will be used in this paper. Readers who are interested in more details about tensors can refer to \cite{Kolda2009,Cichocki2015,Sidiropoulos2017} and the references therein. 
A tensor denotes a multi-dimensional (a.k.a. multi-way or multi-mode) array. The {\it{order}} of a tensor is defined as the number of dimensions of the tensor. A vector and a matrix are special cases of a tensor, i.e., a vector is a tensor of order one, and a matrix is a tensor of order two.

Given an $N$-th order tensor $\cX \in \bWC^{I_1 \times I_2 \times \cdots \times I_N}$, let its $(i_1, i_2, ..., i_N)$-th element be denoted by $x_{i_1 i_2 \cdots i_N}  = \cX(i_1, i_2, \cdots, i_N)$. The {\it{mode-}}$n$ {\it{fibers}} of $\cX$ are defined as  vector-valued sub-tensors obtained by fixing all but one index associated with mode-$n$, i.e., $\cX(i_1, \cdots, i_{n-1}, :, i_{n+1}, \cdots, i_N) $. The number of mode-$n$ fibers in $\cX$ is $\prod_{m=1, m \neq n}^N I_m$ .

The  {\it{mode-}}$n$ {\it{matricization}} (a.k.a.  {\it{unfolding}}) is a process that transforms a tensor  into a matrix whose columns are composed of mode-$n$ fibers of the tensor. The {\it{mode-}}$n$ {\it{unfolding matrix}} of $\cX\in \bWC^{I_1 \times I_2 \times \cdots \times I_N}$ is denoted by $\bX_{(n)} \in \bWC^{I_n \times I_1 I_2  \cdots I_{n-1} I_{n+1} \cdots  I_N} $. The tensor  $\cX(i_1, i_2, \cdots, i_N)$ maps to $\bX_{(n)}(i_n, j)$ such that 
$j = 1 + \sum_{k=1,k \neq n}^N \left(  \left(i_k - 1 \right) \prod_{m=1, m \neq n }^{k-1} I_m \right)$.

The  {\it{mode-}}$n$ {\it{product}} of a tensor $\cX \in \bWC^{I_1 \times I_2 \times \cdots \times I_N}$ and a matrix $\bA \in \bWC^{J \times I_n}$ is denoted by $\cX \times_{n} \bA$. Let $\cY = \cX \times_{n} \bA$. Then, the elements of the tensor $\cY \in \bWC^{I_1 \times I_2 \times \cdots I_{n-1} \times J \times I_{n+1} \times \cdots \times I_N}$ are given by
f
$y_{i_1 i_2 \cdots i_{n-1} j i_{n+1} \cdots i_N} = \sum_{i_n=1}^{I_n} x_{i_1 i_2 \cdots i_{n-1} i_n i_{n+1} \cdots i_N} a_{j i_n}$. The mode-$n$ product representation $\cY = \cX \times_{n} \bA$ can also be expressed by using the mode-$n$ matricization as
\begin{equation}\label{moden_product2}
\bY_{(n)} = \bA \bX_{(n)}. 
\end{equation}

Given a tensor $\cX \in \mathbb{C}^{I_1 \times I_2 \times \cdots \times I_N}$ and matrices $\bA^{(n)} \in \mathbb{C}^{J_n \times I_n}$ for $n=1,...,N$,  their {\it{full multilinear product}}
 is defined as 
\begin{equation}\label{multilinear_product_def}
 \llbracket \cX; \bA^{(1)}, ..., \bA^{(N)} \rrbracket = \cX \times_{1}  \bA^{(1)} \times_{2}  \bA^{(2)} \cdots \times_{N}  \bA^{(N)}.
\end{equation}
For a special case where $I_1 = \cdots = I_N=R$ and $\cX$ is a diagonal tensor $ \cI \in \mathbb{C}^{R \times R \times \cdots \times R}$ that has zero off-diagonal elements and unit diagonal elements, there exists a simplified notation of the full multilinear product as
\begin{equation}\label{multilinear_product_def2}
 \llbracket  \bA^{(1)}, ..., \bA^{(N)} \rrbracket = \cI \times_{1}  \bA^{(1)} \times_{2}  \bA^{(2)} \cdots \times_{N}  \bA^{(N)}. 
\end{equation}

The {\it{norm}} of a tensor is defined as
\begin{equation}\label{tensor_norm_def}
\| \cX \| = \sqrt{\sum_{i_1=1}^{I_1}\sum_{i_1=2}^{I_2} \cdots \sum_{i_N=1}^{I_N} |x_{i_1 i_2 \cdots i_N}|^2},
\end{equation} 
which is analogous to the Frobenius norm in the matrix case. 

Let $\cX \in \mathbb{C}^{I_1 \times I_2 \times \cdots \times I_N}$ and $\cY \in \mathbb{C}^{J_1 \times J_2 \times \cdots \times J_M}$. Then, the {\it{outer product}} (a.k.a. {\it{tensor product}}) of $\cX$ and $\cY$ is denoted by $\cX \circ \cY$. Let $\cZ = \cX \circ \cY$. Then, the elements of the tensor  $\cZ \in  \mathbb{C}^{I_1 \times  \cdots \times I_N \times J_1\times \cdots \times J_M}$ are given by  $z_{i_1 \cdots i_N j_1  \cdots j_M} =  x_{i_1  \cdots i_N} y_{j_1  \cdots j_M}, \forall i_1,...,i_N,j_1,...,j_M$.

A tensor $\cX  \in \mathbb{C}^{I_1 \times I_2 \times \cdots \times I_N }$ is called a {\it{rank-one tensor}} if it can be written as the outer product of vectors as
\begin{equation}\label{rank_one_tensor_def}
\cX = \bx^{(1)} \circ \bx^{(2)} \circ \cdots \circ \bx^{(N)},
\end{equation} 
where $\bx^{(n)} \in \mathbb{C}^{I_n \times 1}, \forall n$.

The canonical polyadic decomposition (CPD), which is also known as CANDECOMP/PARAFAC decomposition,  factorizes a tensor into a sum of component rank-one tensors. The CPD of $\cX \in  \mathbb{C}^{I_1 \times I_2 \times \cdots \times I_N } $ has a form 
\begin{equation}\label{CPD_form}
\cX = \sum_{r=1}^R \bx_r^{(1)} \circ \bx_r^{(2)}  \circ \cdots \circ \bx_r^{(N)},
\end{equation} 
where $\bx_r^{(n)} \in \mathbb{C}^{I_n \times 1 }$ for $r=1,...,R$. The minimum possible value of the number of rank-one tensors that constitute $\cX$, which is $R$ in \eqref{CPD_form},  is called the {\it{rank}} of   $\cX$.

\section{Channel model and system model}\label{sec:sys_model}

Consider a TDD system where a base station with $\Nt$ antennas and $\Nrf (\leq \Nt)$ RF chains communicates with a mobile station that has a single antenna. 

\subsection{Channel model}\label{subsec:ch_model}

We consider a spatially sparse channel that has $\Lch$ paths between the BS and mobile user. 
Let $\tau_\ell$ and $\phi_{\ell}$ denote the path delay and AoA of the $\ell^{\mathrm{th}}$  path. Let $g_{t,\ell}$ denote the short-term fading complex path gain of the $\ell^{\mathrm{th}}$ path at frame $t$. 
Let $p_{\mathrm{PS}}(\tau)$ denote the low pass filter including pulse shaping and analog filters. 
We assume a uniform linear array (ULA) with antenna element spacing $d_{\mathrm{a}}$ and signal wavelength $\lambda$. 
It is possible to extend the proposed method to a uniform planar array (UPA).
The array response vector  associated with the $\ell^{\mathrm{th}}$ AoA $\phi_{\ell}$  is expressed as
\begin{equation}\label{ULA_def}
\begin{split}
 \ba(\phi_{\ell}) = \begin{bmatrix} 1 &  e^{  \frac{j 2 \pi d_{\mathrm{a}} \sin(\phi_{\ell})}{\lambda} } & \cdots &  e^{  \frac{j 2 \pi d_{\mathrm{a}} (\Nt-1) \sin(\phi_{\ell})}{\lambda} } \end{bmatrix}^{\mathsf{T}}.
\end{split}
\end{equation}
Let $T_s$ and $N_{\mathrm{CP}}$ be the sampling duration and the cyclic prefix length. We assume that the large-scale fading parameters, $\tau_\ell$'s and $\phi_{\ell}$'s,  are constant during the estimation process.
By using the delay-$d$ channel model \cite{SchniterSayeed2014,Alkhateeb2015a,WINNER2}, the uplink channel at frame $t$ can be represented as
\begin{equation}\label{WB_ch_model_CIR}
\bh_{t}[d]  = \sum_{\ell=1}^{\Lch} g_{t,\ell} p_{\mathrm{PS}}(dT_s - \tau_{\ell}) \ba ( \phi_{\ell}) \;\; \text{for} \;\; d=0,...,  N_{\mathrm{CP}}-1. 
\end{equation}
By letting $c_{k,\ell} = \sum_{d=0}^{N_{\mathrm{CP}}-1} p_{\mathrm{PS}}(dT_s - \tau_{\ell}) e^{ -\frac{j 2 \pi (k-1) d}{\Kf}}$, the channel frequency response vector  can be expressed as
\begin{equation}\label{WB_ch_model_CFR}
\begin{split}
\bh_{t,k}  
&=\sum_{\ell=1}^{\Lch} g_{t,\ell} c_{k,\ell} \ba( \phi_{\ell}),  
\end{split}
\end{equation}
at frame $t$ and subcarrier $k$.

\subsection{System model}\label{subsec:sys_model}

Let $s_{t,k}$ be an uplink training symbol at frame $t$ and subcarrier $k$ with $|s_{t,k}|=1$, and $\bz_{t,k} \sim \mathcal{CN}(\mathbf{0}, \sigma^2 \bI)$ be a circularly symmetric Gaussian noise.  The $\Nt \times 1$ received signal vector at each frame and subcarrier can be represented as 
\begin{equation}\label{rx_signal}
\begin{split}
\br_{t,k} = \bh_{t,k} s_{t,k} + \bz_{t,k}.
\end{split}
\end{equation}

Let $\bW_{\mathrm{RF}} \in \mathbb{C}^{\Nt \times \Nrf}$ and $\bW_{\mathrm{BB}} \in \mathbb{C}^{\Nrf \times \Nrf}$ be an analog combining matrix and  a digital baseband processing matrix.
Similar to a sensing matrix in prior CS-based channel estimation work \cite{Rodriguez2018twc,Venugopal2017jsac}, 
we assume that the elements of $\bW_{\mathrm{RF}}$ have random phases with a unit amplitude. 
Let $\bW$ denote the hybrid combining matrix as $\bW =  \bW_{\mathrm{RF}} \bW_{\mathrm{BB}}  $. After combining with the hybrid combiner and multiplying by $s_{t,k}^*$, the $\Nrf \times 1$ baseband received signal vector becomes  
\begin{equation}\label{rx_signal_vector_tk}
\begin{split}
\by_{t,k} = s_{t,k}^* \bW^* \br_{t,k} = \bx_{t,k} +  \bn_{t,k},
\end{split}
\end{equation}
where $\bx_{t,k} = \bW^* \bh_{t,k}$ denotes the signal part of $\by_{t,k}$, and $\bn_{t,k} = s_{t,k}^* \bW^* \bz_{t,k} \sim \mathcal{CN}(\mathbf{0}, \sigma^2 \bW^* \bW) $ denotes the noise part. From the viewpoint of the estimator at baseband, the effective noise $\bn_{t,k}$ becomes colored for an arbitrary hybrid combiner $\bW$. It is possible to whiten the effective noise by using the baseband combiner $\bW_{\mathrm{BB}} = \left(\bW_{\mathrm{RF}}^* \bW_{\mathrm{RF}} \right)^{-\frac{1}{2}}$. With this choice of $\bW_{\mathrm{BB}}$, the hybrid combiner $\bW$ satisfies  $\bW^* \bW = \bI $ for any $\bW_{\mathrm{RF}}$. We will use this unitary hybrid combiner $\bW$ throughout this paper.


\subsection{Tensor representation of channels and received signals}\label{subsec:tensor_model}

In this subsection, we show that the time-varying frequency-selective channel  in \eqref{WB_ch_model_CFR} can be represented as a low-rank third-order tensor.  
Let $\ba_{\ell} = \ba( \phi_{\ell})  $, $\bc_{\ell} = \begin{bmatrix} c_{1,\ell} & \cdots & c_{\Kf,\ell} \end{bmatrix}^{\mathsf{T}} $, and $\bg_{\ell} = \begin{bmatrix} g_{1,\ell} & \cdots & g_{\Tt,\ell} \end{bmatrix}^{\mathsf{T}} $  for $\ell =1,...,\Lch$. 
With $\ba_{\ell}$, $\bc_{\ell}$, and $\bg_{\ell}$,  let us define $\bA$, $\bC$, and $\bG$ as   $\bA =\begin{bmatrix} \ba_1 & \cdots &  \ba_\Lch \end{bmatrix}$, $\bC  =\begin{bmatrix} \bc_1 & \cdots & \bc_\Lch \end{bmatrix} $, and  $\bG =\begin{bmatrix} \bg_1 & \cdots & \bg_\Lch \end{bmatrix}  $.

\begin{figure}[!t]
	\centerline{\resizebox{0.68\columnwidth}{!}{\includegraphics{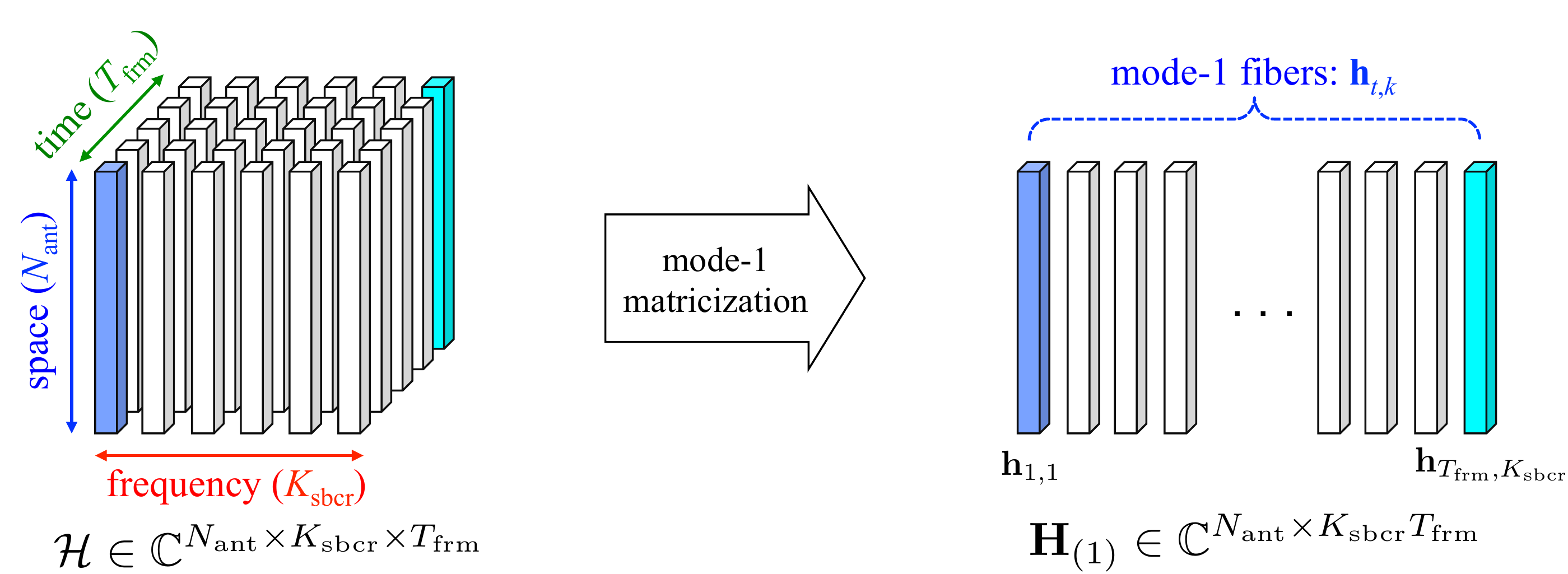}}}
	\caption{Tensor representation of a time-varying frequency-selective channel: the left figure shows the third-order tensor representation $\cH \in \mathbb{C}^{\Nt \times \Kf \times \Tt}$, and the right figure shows its mode-1 unfolding matrix $\bH_{(1)} \in \mathbb{C}^{\Nt \times \Kf \Tt}$. }
	\label{fig:channel_model_3D_tensor}
\end{figure}

The $\Kf \Tt $ channel frequency response vectors $\bh_{t,k} \in \mathbb{C}^{\Nt \times 1}$ for $t=1,...,\Tt$ and $k=1,...\Kf$ in \eqref{WB_ch_model_CFR} can be regarded as $\Tt \Kf$ mode-1 fibers in a third-order tensor $\cH  \in \mathbb{C}^{\Nt \times \Kf \times \Tt}$ as shown in \figref{fig:channel_model_3D_tensor}. In CPD form, the rank-$\Lch$ third-order tensor is
\begin{equation}\label{def_3D_H}
\begin{split}
\mathcal{H} &= \llbracket   \bA, \bC, \bG \rrbracket \\
&= \mathcal{I} \times_1 \bA \times_2 \bC \times_3 \bG\\
&=\sum_{\ell=1}^{\Lch} \ba_{\ell} \circ \bc_{\ell} \circ \bg_{\ell}.
\end{split}
\end{equation}
In this CPD form in \eqref{def_3D_H}, the channel tensor $\cH$ is factorized into the three matrices, $\bA$, $\bC$ and $\bG$, which are called factor matrices. The mode-1 factor matrix $\bA$ is associated with antennas in the space domain, the mode-2 factor matrix $\bC$ with subcarriers in the frequency domain, and the mode-3 factor matrix $\bG$ with frames in the time domain.

From the channel frequency response tensor model, the received signal at baseband, $\by_{t,k}, \forall t,k$ in \eqref{rx_signal_vector_tk} can also be represented as a third-order tensor as
\begin{equation}\label{def_3D_Y}
\begin{split}
\cY &= \cX + \cN, 
\end{split}
\end{equation}
where $\cN$ is the noise tensor whose mode-1 fibers are IID Gaussian vectors with $\mathcal{CN}(\mathbf{0}, \sigma^2 \bI) $. The signal tensor $\cX$ is given by 
\begin{equation}\label{def_3D_X}
\begin{split}
\cX=\cH  \times_1 \bW^* \in \mathbb{C}^{\Nrf \times \Kf \times \Tt} .  
\end{split}
\end{equation}
The tensor representation in \eqref{def_3D_X} can also be expressed by using a mode-1 matricization as $\bX_{(1)} =  \bW^* \bH_{(1)}$
where $\bH_{(1)} = \bA \left( \bG \odot \bC \right)^{\mathsf{T}}$ is the mode-1 matricization of $\cH$ shown in \figref{fig:channel_model_3D_tensor}.
Let $\bB = \begin{bmatrix} \bb_1 & \cdots & \bb_\Lch \end{bmatrix} = \bW^* \bA \in \mathbb{C}^{\Nrf \times \Lch}$, which can be regarded as the effective array response matrices from the viewpoint of the baseband estimator. 
The CPD of $\cX$ is given by 
\begin{equation}\label{def_3D_X_CPD}
\begin{split}
\cX &= \llbracket  \bB, \bC, \bG \rrbracket.
\end{split}
\end{equation}
Note that both the time-varying frequency-selective channel $\cH$ and its associated baseband received signal part $\cX$ are  third-order tensors with rank $\Lch$.

\section{Spatial channel covariance estimation based on tensor decomposition}\label{sec:Cov_est_tensor}

The spatial channel covariance matrix can be estimated from 
the sample covariance of $\Kf \Tt $ mode-1 fibers in $\cH$. Since we assume that the elements in $\cH$ have zero mean, the sample covariance of the mode-1 fibers  becomes
\begin{equation}\label{sample_spatial_ch_cov_ideal}
\begin{split}
\bR_{\bh} &= \frac{1}{\Kf \Tt} \sum_{k=1}^\Kf \sum_{t=1}^\Tt \cH(:,k,t) \cH(:,k,t)^*\\
&= \frac{1}{\Kf \Tt} \bH_{(1)} \bH_{(1)}^* \\
&=\frac{1}{\Kf \Tt} \bA \left( \bG \odot \bC \right)^{\mathsf{T}} \left( \bG \odot \bC \right)^{\mathsf{C}} \bA^*\\
&= \frac{1}{\Kf \Tt} \bA \left( \bG^* \bG  \circledcirc  \bC^* \bC \right) \bA^*.
\end{split}
\end{equation}
The goal of the spatial channel covariance estimation for the hybrid architecture is  to calculate $\bR_{\bh}$ of $\cH \in \mathbb{C}^{\Nt \times \Kf \times  \Tt} $   from the baseband received signal   $\cY  \in \mathbb{C}^{\Nrf \times \Kf \times  \Tt} $, which has smaller dimensions than $\cH$. In this section, we propose an estimation method that has three steps. 
In the first step, the measurement tensor $\cY$ is decomposed into three factor matrices: $\hat{\bB}$, $\hat{\bC}$, and $\hat{\bG}$  
in a CPD form. 
Note that the  three factor matrices obtained in the first step are strongly related to the actual factor matrices $\bB$, $\bC$, and $\bG$, but are not identical. 
The relationship between these matrices is further explained in \sref{Proposed_step1}. 
In the second step, the estimate of the actual factor matrix $\bA$ of the channel tensor $\cH$ is obtained from $\hat{\bB}$, which is denoted by $\hat{\bA}$. The spatial channel covariance matrix is calculated from $\hat{\bA}$, $\hat{\bC}$, and $\hat{\bG}$ in the last step. Each step is explained in detail in the following subsections. 

\subsection{First step: factorization of $\cY$ in a CPD form}\label{Proposed_step1}

If the factor matrices $\bB$, $\bC$, and $\bG$ are given, we can exactly calculate the signal part $\cX$ of the measurement tensor $\cY$. 
The reverse process does not hold in general; 
the perfect reconstruction of the original $\bB$, $\bC$, and $\bG$ from any $\cX$ are not guaranteed.   
There is, however, a special case where the factor matrices can be reconstructed from  $\cX$. 
If the tensor rank of a higher-order (more than second-order) is low, its CPD is unique under some mild constraints \cite{Cichocki2015}. The uniqueness of the CPD means that there exists  only one possible combination of rank-one tensors that sum to the given tensor subject to two types of indeterminacy: scaling and permutation. 
The scaling indeterminacy means that the columns in each factor matrix can be scaled arbitrarily, i.e., the CPD form in \eqref{def_3D_X_CPD} can be rewritten as
 \begin{equation}\label{scaling_indermin_vector}
\begin{split}
\cX &= \sum_{\ell=1}^{\Lch} (\delta_{\bb,\ell} \bb_{\ell}) \circ (\delta_{\bc,\ell}\bc_{\ell})  \circ (\delta_{\bg,\ell} \bg_{\ell}), 
\end{split}
\end{equation}
as long as $\delta_{\bb,\ell} \delta_{\bc,\ell} \delta_{\bg,\ell}= 1$ for $\ell =1, ..., \Lch$. The CPD form in \eqref{scaling_indermin_vector} can be expressed in a multilinear product format as
\begin{equation}\label{scaling_indermin_matrix}
\begin{split}
\cX &= \left \llbracket \bB \mathbf{\Delta}_{\bB}, \bC \mathbf{\Delta}_{\bC}, \bG \mathbf{\Delta}_{\bG} \right \rrbracket,
\end{split}
\end{equation}
where $\mathbf{\Delta}_{\bB}$, $\mathbf{\Delta}_{\bC}$, and $\mathbf{\Delta}_{\bG}$ are any diagonal matrices satisfying $\mathbf{\Delta}_{\bB} \mathbf{\Delta}_{\bC} \mathbf{\Delta}_{\bG} = \bI$.
The permutation indeterminacy means that the column vectors in each factor matrix can be reordered with a permutation matrix that is common all the modes, i.e.,  the CPD form in \eqref{def_3D_X_CPD} can also be represented as
\begin{equation}\label{permutation_indermin}
\begin{split}
\cX &= \left \llbracket \bB \mathbf{\Pi}, \bC \mathbf{\Pi}, \bG \mathbf{\Pi} \right \rrbracket ,
\end{split}
\end{equation} 
for any 
permutation matrix $ \mathbf{\Pi}$. From \eqref{scaling_indermin_matrix} and \eqref{permutation_indermin}, the general form of the CPD of  $\cX$ becomes
\begin{equation}\label{CPD_wIndeterminacy}
\begin{split}
\cX &= \left \llbracket \bB \mathbf{\Pi}  \mathbf{\Delta}_{\bB}  , \bC \mathbf{\Pi} \mathbf{\Delta}_{\bC}  , \bG \mathbf{\Pi} \mathbf{\Delta}_{\bG} \right \rrbracket.
\end{split}\
\end{equation} 

If we set the indeterminacy issue aside, we can exactly reconstruct its factor matrices by leveraging the uniqueness of the CPD. 
In this subsection, we focus on how to find a CPD solution. We will discuss how to deal with the indeterminacy in the following subsections.

Given a  received signal tensor  $\cY$, the problem of finding its CPD form is formulated as
\begin{equation}\label{ALS_prob}
\begin{split}
\{ \hat{\bB},\hat{\bC},\hat{\bG} \} =\mathop{\mathrm{arg \, min}}_{\mB,\mC,\mG} \left \| \cY - \left \llbracket \mB, \mC, \mG \right \rrbracket \right \|. 
\end{split}
\end{equation}
There are many known algorithms to solve \eqref{ALS_prob} for CPD.  
One algorithm to compute the CPD problem in \eqref{ALS_prob} is the alternating least squares (ALS) \cite{Kolda2009}. 
By rewriting the objective function in \eqref{ALS_prob} in matrix form as 
\begin{equation}\label{ALS_prob1}
\begin{split}
 \left \| \cY - \left \llbracket \mB, \mC, \mG \right \rrbracket \right \| 
&= \| \bY_{(1)} - \mB (\mG \odot \mC)^{\mathsf{T}} \|_F \\
&= \| \bY_{(2)} - \mC (\mG \odot \mB)^{\mathsf{T}} \|_F \\
&= \| \bY_{(3)} - \mG (\mC \odot \mB)^{\mathsf{T}} \|_F,
\end{split}
\end{equation}
 the ALS algorithm first  finds the mode-1 factor matrix $\mB$ assuming that the mode-2 and mode-3 factor matrices, $\mC$ and $\mG$ are fixed. This subproblem is formulated as 
\begin{equation}\label{ALS_subprob_mode1}
\begin{split}
\min_{\mB} \| \bY_{(1)} - \mB (\mG \odot \mC)^{\mathsf{T}} \|_F.  
\end{split}
\end{equation}
The solution to \eqref{ALS_subprob_mode1} can be found by using the least squares algorithm as
\begin{equation}\label{ALS_subprob_mode1_sol}
\begin{split}
\mB &=  \bY_{(1)} \left(\left (\mG \odot \mC \right)^{\mathsf{T}} \right)^{\dagger} \\
&=  \bY_{(1)} \left( \left (\mG \odot \mC \right) \left( \mG^* \mG \circledcirc \mC^* \mC  \right)^{\dagger} \right)^{\mathsf{C}}. 
\end{split}
\end{equation}

Similar to  \eqref{ALS_subprob_mode1}, the mode-2 factor matrix $\mC$ and the mode-3 factor matrix $\mG$ can  be calculated by fixing other factor matrices except for its own factor matrix. 
The ALS algorithm iterates the three steps until the objective function converges. 
The convergence is guaranteed although the converged solution may not be a global optimum.

The solution after convergence provides an estimate of the CPD form of $\cX$ as
\begin{equation}\label{CPD_ALS_solution}
\begin{split}
\hat{\cX} &= \left \llbracket \hat{\bB}, \hat{\bC}, \hat{\bG} \right \rrbracket.
\end{split}
\end{equation} 
Note that $\hat{\cX}$ is an estimate of the actual $\cX$ in \eqref{CPD_wIndeterminacy}.
Due to the scaling and permutation indeterminacy, the estimated factor matrices $\hat{\bB}$, $\hat{\bC}$, and $\hat{\bG}$ obtained in the first step are related to the actual factor matrices $\bB$, $\bC$, and $\bG$ as
\begin{equation}\label{relation_fac_mat}
\begin{split}
\hat{\bB} &= \bB \mathbf{\Pi} \mathbf{\Delta}_{\bB}  + \mathbf{\Omega}_{\bB},\\
\hat{\bC} &= \bC \mathbf{\Pi} \mathbf{\Delta}_{\bC} + \mathbf{\Omega}_{\bC},\\
\hat{\bG} &=\bG \mathbf{\Pi} \mathbf{\Delta}_{\bG}  + \mathbf{\Omega}_{\bG}, 
\end{split}
\end{equation}
where $\mathbf{\Delta}_{\bB}$, $\mathbf{\Delta}_{\bC}$, and $\mathbf{\Delta}_{\bG}$ are  complex-valued diagonal matrices that satisfy $\mathbf{\Delta}_{\bB} \mathbf{\Delta}_{\bC} \mathbf{\Delta}_{\bG} = \bI$, and $\mathbf{\Omega}_{\bB}$, $\mathbf{\Omega}_{\bC}$, and $\mathbf{\Omega}_{\bG}$ denote the estimation errors caused by CPD.

\subsection{Second step: estimation of $\bA$ of the channel tensor $\cH$}\label{Proposed_step2}

The goal of the second step is to estimate $\bA \mathbf{\Pi}$ from  $\hat{\bB}$ that is obtained in the first step as described in \sref{Proposed_step1}. Let $\breve{\bA} =  \bA \mathbf{\Pi}$. We will show in \sref{Proposed_step3} that we do not need to  obtain $\mathbf{\Pi}$ explicitly to estimate the spatial channel covariance.
Let  $\ba(\hat{\phi}_{\ell})$ denote  the $\ell$-th column in the estimate of $\breve{\bA}$. 
The problem of finding $\hat{\phi}_{\ell}$ that minimizes the angle between  $\hat{\bb}_{\ell}$ and $\bW^*\ba(\hat{\phi}_{\ell})$ is represented as 
\begin{equation}\label{prob_find_AoA}
\begin{split}
\hat{\phi}_{{\ell}} &= \arg \min_{\phi} \left( 1- \frac{ | \hat{\bb}_{\ell}^* \bW^* \ba(\phi) |^2 }{\|\hat{\bb}_{\ell} \|^2 \| \bW^* \ba(\phi) \|^2 } \right),
\end{split}
\end{equation}
and its solution can be found by one-dimensional search methods with respect to $\phi$, which is a continuous variable. Since we assume ULA, the solution can be obtained more efficiently by using a polynomial equation similar to the Root-MUSIC algorithm \cite{Rao1989tassp}. 
By letting $z = e^{\frac{j 2 \pi d_{\mathrm{a}} \sin(\phi)}{\lambda}} $, the array response vector $\ba(\phi)$ can be denoted by $\ba(z) = \begin{bmatrix} 1 & z & \cdots & z^{\Nt-1} \end{bmatrix}^T$.
Then, the optimization problem in \eqref{prob_find_AoA} is rewritten as 
\begin{equation}\label{prob_find_AoA_z}
\begin{split}
\hat{z}_{{\ell}} 
 &=  \arg \min_{z} \left(  \frac{ \ba^*(z) \bW \left( \|\hat{\bb}_{\ell} \|^2 \bI  - \hat{\bb}_{\ell} \hat{\bb}_{\ell}^*  \right) \bW^* \ba(z) }{\|\hat{\bb}_{\ell} \|^2 \ba^*(z) \bW \bW^* \ba(z)  } \right).
\end{split}
\end{equation}
Let $\bQ_{\ell} = \bW \left( \|\hat{\bb}_{\ell} \|^2 \bI  - \hat{\bb}_{\ell} \hat{\bb}_{\ell}^*  \right) \bW^*$. The numerator in \eqref{prob_find_AoA_z} is represented as a polynomial with respect to $z$ and becomes zero in the noiseless case, i.e.,
\begin{equation}\label{prob_find_AoA_poly_eq}
\begin{split}
\ba^*(z) \bQ_{\ell} \ba(z) 
&= \sum_{m=-\Nt+1}^{\Nt-1} \left( \sum_{n_1 - n_2 = m}  [\bQ_{\ell}]_{n_1,n_2} \right) z^m =0.
\end{split}
\end{equation}
Note that if $\omega$ is a root of \eqref{prob_find_AoA_poly_eq}, then $1/\omega^*$ is also its root, and there are $(\Nt-1)$ roots within the unit circle. Let $\omega_1, ..., \omega_{\Nt-1}$ denote the  $(\Nt-1)$ roots normalized by their amplitudes. 
Then, the solution to \eqref{prob_find_AoA_z} can be obtained by  searching over  $z \in \{ \omega_1, ..., \omega_{\Nt-1} \}$ that has discrete $(\Nt-1)$ elements, i.e., $\hat{z}_{{\ell}} = \arg \max_{z \in \{ \omega_1, ..., \omega_{\Nt-1} \}}  \frac{ | \hat{\bb}_{\ell}^* \bW^* \ba(z) |^2 }{ \| \bW^* \ba(z) \|^2 } $.
 After obtaining $\hat{z}_{\ell}$, 
the diagonal elements in $\mathbf{\Delta}_{\bB}$ can be estimated as
\begin{equation}\label{solution_lambda}
\begin{split}
\hat{\delta}_{\bB,\ell} &= \frac{\ba^*(\hat{z}_{\ell} ) \bW \hat{\bb}_{\ell} }{\|\bW^* \ba(\hat{z}_{\ell} ) \|^2}.
\end{split}
\end{equation}



Let $\hat{\mathbf{\Delta}}_{\bB}$ and  $\hat{\bA}$ denote the estimate of $\mathbf{\Delta}_{\bB}$ and $\breve{\bA}$. Then,  $\hat{\mathbf{\Delta}}_{\bB}$ and $\hat{\bA}$ can be represented as    $\hat{\mathbf{\Delta}}_{\bB} = \mathrm{diag} \left( \begin{bmatrix} \hat{\delta}_{\bB,1} & \cdots &\hat{\delta}_{\bB,\Lch}   \end{bmatrix} \right)$ and   $\hat{\bA} = \begin{bmatrix} \ba(\hat{z}_1) & \cdots &\ba(\hat{z}_\Lch)  \end{bmatrix}$. Note that $\hat{\bA}$ is the estimate of $\breve{\bA} = \bA \mathbf{\Pi}$, in which the permutation matrix $\mathbf{\Pi}$ is not known. 


\subsection{Third step: estimation of the spatial channel covariance matrix from $\tilde{\bA}$, $\hat{\bC}$, and $\hat{\bG}$}\label{Proposed_step3}


While the estimate of $\bA$ is given by $\hat{\bA}$ with only permutation indeterminacy at the second step in \sref{Proposed_step2}, the estimated factor matrices $\hat{\bG}$ and $\hat{\bC}$ at the first step in \sref{Proposed_step1} still contain both  scaling and permutation indeterminacy. Consequently, it is impossible to simply replace $\bG$ and $\bC$ by $\hat{\bG}$ and  $\hat{\bC}$ in \eqref{sample_spatial_ch_cov_ideal} without considering $\mathbf{\Delta}_{\bC}$, $\mathbf{\Delta}_{\bG}$, and $ \mathbf{\Pi}$.

Let $\tilde{\bA}$, $\tilde{\bC}$ and $\tilde{\bG}$ denote the estimate of the actual $\bA$, $\bC$ and $\bG$ without any indeterminacy, which are defined as $\tilde{\bA} = \hat{\bA} \mathbf{\Pi}^*$, $\tilde{\bC}    = \hat{\bC}  \mathbf{\Delta}_{\bC}^{-1} \mathbf{\Pi}^* $ and $
\tilde{\bG}  =  \hat{\bG} \mathbf{\Delta}_{\bG}^{-1} \mathbf{\Pi}^* $. 
Note that $\mathbf{\Pi}^{-1} =  \mathbf{\Pi}^*$ for any permutation matrix $\mathbf{\Pi}$.
Then, the estimate of the sample spatial channel covariance matrix in \eqref{sample_spatial_ch_cov_ideal} can be calculated from $\tilde{\bA}$, $\tilde{\bC}$, and $\tilde{\bG}$ as
\begin{equation}\label{sample_spatial_ch_cov_est}
\begin{split}
\tilde{\bR}_{\bh} 
&= \frac{1}{\Kf \Tt} \tilde{\bA} \left( \tilde{\bG}^* \tilde{\bG}  \circledcirc  \tilde{\bC}^* \tilde{\bC} \right) \tilde{\bA}^*\\
&\stackrel{(a)}{\approx} \frac{1}{\Kf \Tt} \hat{\bA} \hat{\mathbf{\Delta}}_{\bB}^*\left( \hat{\bG}^* \hat{\bG} \circledcirc   \hat{\bC}^* \hat{\bC} \right) \hat{\mathbf{\Delta}}_{\bB} \hat{\bA}^*,
\end{split}
\end{equation}
where $(a)$ comes from  the fact that $\mathbf{\Delta}_{\bB} \mathbf{\Delta}_{\bC} \mathbf{\Delta}_{\bG} = \bI$, and $\hat{\mathbf{\Delta}}_{\bB}$ is the estimate of $\mathbf{\Delta}_{\bB}$.
\section{Relative precoding efficiency for spatial channel covariance estimation}\label{sec:Performance_metric}

The MSE or normalized MSE (NMSE) is typically used as a performance metric for channel estimation methods.
Other metrics, though, are more relevant for spatial channel covariance estimation. 
This is because the dominant eigenvalues and their eigenspaces are more useful for hybrid precoding rather than each element in the covariance matrix. 
In this regard, we adopt the performance metric used in \cite{DBLP:journals/corr/HaghighatshoarC15::9,Park2017twc_Cov}, which we call relative precoding efficiency (RPE).  
Let $\bR_{\bh} \in \mathbb{C}^{\Nt \times \Nt}$ and $\tilde{\bR}_{\bh} \in \mathbb{C}^{\Nt \times \Nt} $ be the spatial channel covariance and its estimate.
Let $\bU$ and $\tilde{\bU}$ denote the matrices composed of the dominant $\Nrf$ eigenvectors of $\bR_{\bh}$ and $\tilde{\bR}_{\bh}$.
The RPE is defined as
\begin{equation} \label{eff_metric_def}
\eta = \frac{\Tr(\tilde{\bU}^* \bR_{\bh} \tilde{\bU} )}{ \Tr(\bU^* \bR_{\bh} \bU )}. 
\end{equation} 
This metric lies between  zero and one, i.e., $0 \leq \eta \leq 1$, and  higher $\eta$ indicates  more accurate estimation. 
The RPE $\eta$ in \eqref{eff_metric_def} is closely related to the relative spectral efficiency of the hybrid beamforming based on $\tilde{\bR}_{\bh}$   compared to that of the hybrid beamforming based on $\bR_{\bh}$. 
Consider a hybrid beamforming system where the analog part is composed of $ \bU $ or $ \tilde{\bU} $ with $\Nrf$ RF chains as in \cite{Park2017twc, Park2017tsp}. For analytical tractability, we ignore the fact that phase shifter are typically used for the analog part.  
At low SNR region, the achievable rate ratio approximates to
\begin{equation}
\begin{split}
\frac{\mathrm{rate^{(est.)}_{hybrid}}}{\mathrm{rate^{(ideal)}_{hybrid}}} 
&= \frac{\mathbb{E}\left[ \log (1 + \frac{1}{\sigma^2}\bh^* \tilde{\bU}  \tilde{\bU}^* \bh)   \right]}{\mathbb{E}\left[ \log (1 + \frac{1}{\sigma^2}\bh^* \bU \bU^* \bh) \right]}   \\
&\stackrel{(a)}{\approx} \frac{\mathbb{E}\left[ \frac{1}{\sigma^2}\bh^* \tilde{\bU}  \tilde{\bU}^* \bh  \right]}{\mathbb{E}\left[  \frac{1}{\sigma^2}\bh^* \bU \bU^* \bh \right]}   \\
&\stackrel{(b)}{=}  \frac{ \mathrm{Tr}(  \tilde{\bU}^* \bR_{\bh} \tilde{\bU} ) }{\mathrm{Tr}( \bU^* \bR_{\bh} \bU  )} ,
\end{split}
\end{equation}
where $(a)$ comes from the fact that $\ln(1+x) \approx x$ for  $x \approx 0$, and $(b)$ comes from  $\mathrm{Tr}(\bA \bB) = \mathrm{Tr}(\bB \bA)$ and $\bR_{\bh} = \mathbb{E}[ \bh \bh^*] $. This result shows that the metric $\eta$ allows us to anticipate how much relative loss will be caused by the estimation error in terms of achievable rate  at  low SNR region.  

The RPE $\eta$ defined in \eqref{eff_metric_def} can be analyzed approximately in  large antenna array regimes.  
Let  $\bA \in \mathbb{C}^{\Nt \times \Lch}$ be a matrix composed of array response vectors and $\bR_{\bg} = \mathbb{E}[\bg \bg^*] \in \mathbb{C}^{\Lch \times \Lch}$ be the covariance of channel path gains. The spatial channel covariance matrix is represented as
\begin{equation} \label{R_def}
\begin{split}
\bR_{\bh} = \bA \bR_{\bg} \bA^*.
\end{split}
\end{equation}
As $\Nt$ becomes large, $\bA^* \bA \approx \Nt \bI$, i.e., $\frac{1}{\sqrt{\Nt}} \bA$ becomes semi-unitary asymptotically \cite{Ayach2012SPAWC}. 
If we assume that $g_{t,k}$ are IID complex random variables with zero mean and variance $1/\Lch$ 
for analytical tractability, then \eqref{R_def} 
can be regarded as the approximate eigenvalue decomposition  of $\bR_{\bh}$, i.e., $\bU \approx \frac{1}{\sqrt{\Nt}}\bA$.
Let $e_{\ell}$ be the AoA estimation error and $\tilde{\phi}_{\ell} = \phi_{\ell} + e_{\ell} $  be the estimated AoA for the $\ell$-th path. Similar to $\bU$, we assume that $\tilde{\bU}$  approximates to  $\frac{1}{\sqrt{\Nt}}\tilde{\bA}$ as $\Nt$ increases. 
Then, the RPE $\eta$ becomes
\begin{equation} \label{eff_metric_1}
\begin{split}
\eta 
&\approx \frac{\Tr \left(\tilde{\bU}^* \bU \left( \frac{\Nt}{\Lch} \bI_{\Lch} \right) \bU^* \tilde{\bU} \right)}{ \Tr \left(  \frac{\Nt}{\Lch} \bI_{\Lch} \right)} \\
&= \frac{\Tr(\tilde{\bU}^* \bU \bU^* \tilde{\bU})}{ \Lch} \\
&\approx \frac{1}{\NtS   \Lch} \| \bA^* \tilde{\bA} \|^2_{\mathrm{F}} \\
& \approx \frac{1}{\NtS   \Lch}   \sum_{\ell=1}^{\Lch} |\ba^*(\phi_{\ell}) \ba(\phi_{\ell} + e_{\ell}) |^2,
\end{split}
\end{equation}
where we assume that $e_{\ell}$ is small and $\frac{1}{N} \ba^*(\phi_{\ell_1}) \ba(\phi_{\ell_2}) \approx 0$ for $\ell_1 \neq \ell_2$.

Let $\kappa_{\ell} $ be defined as
\begin{equation} \label{def_kappa_def}
\begin{split} 
\kappa_{\ell} &= \frac{  \pi d_{\mathrm{a}} }{\lambda} \left(\sin \left(\phi_{\ell} + e_{\ell}) - \sin(\phi_{\ell} \right) \right)\\
&= \frac{  \pi d_{\mathrm{a}} }{\lambda} \left(\sin (\phi_{\ell})(\cos(e_{\ell})-1) + \cos(\phi_{\ell}) \sin(e_{\ell}) \right),
\end{split}
\end{equation}
which approximates to 
$\kappa_{\ell} \approx  \frac{  \pi d_{\mathrm{a}} }{\lambda} \cos(\phi_{\ell}) e_{\ell} $ for small $e_{\ell}$.
In the ULA case, $\eta $ in \eqref{eff_metric_1} is given by
\begin{equation} \label{eff_metric_2}
\begin{split}
\eta & \approx \frac{1}{\NtS   \Lch}   \sum_{\ell=1}^{\Lch} |\ba^*(\phi_{\ell}) \ba(\phi_{\ell} + e_{\ell}) |^2 \\
&=  \frac{1}{\NtS   \Lch}   \sum_{\ell=1}^{\Lch}  \frac{\sin^2 \left( \Nt \kappa_{\ell} \right)}{ \sin^2 \left(  \kappa_{\ell} \right) } \\
& \stackrel{(a)}{\approx}    \frac{1}{ \Lch}  \sum_{\ell=1}^{\Lch}  \left( 1 - \frac{\NtS    \kappa_{\ell}^2}{3}    \right) \\
& \approx   \frac{1}{ \Lch}  \sum_{\ell=1}^{\Lch}  \left( 1 - \frac{\NtS   \pi^2 d_{\mathrm{a}}^2 \cos^2(\phi_{\ell})  }{3 \lambda^2} e_{\ell}^2   \right), 
\end{split}
\end{equation}
where $(a)$ comes from the second-order approximation of Maclaurin series for small $\kappa_{\ell}$.
Consequently, $1-\mathbb{E}[\eta]$  approximately becomes
\begin{equation} \label{eff_metric_mean}
\begin{split}
1 - \mathbb{E} [\eta] 
& \approx   \frac{\NtS   \pi^2 d_{\mathrm{a}}^2 }{3 \Lch \lambda^2}    \sum_{\ell=1}^{\Lch}  \cos^2(\phi_{\ell})  \mathbb{E} \left[(\phi_{\ell} - \tilde{\phi}_{\ell})^2 \right]    \\
& \geq     \frac{\NtS   \pi^2 d_{\mathrm{a}}^2  }{3 \Lch \lambda^2} \sum_{\ell=1}^{\Lch}    \cos^2(\phi_{\ell}) \mathrm{CRLB}(\phi_{\ell}),  
\end{split}
\end{equation}
where the CRLB of the $\phi_{\ell}$ estimation $\mathrm{CRLB}(\phi_{\ell})$ will be derived in the following section. 


\section{Cram\'er-Rao Lower Bound for the AoA estimation}\label{sec:CRLB}

In this section, we derive the CRLB of the MSE of AoAs. 
The basic tool to derive CRLB is based on the method in \cite{LiuSidiropoulos2001} and \cite{Zhou2017}. In \cite{LiuSidiropoulos2001}, all the factor matrices are non-structure matrices, i.e., a factor matrix $\bF \in \mathbb{C}^{\Nt \times M}$ is determined by $\Nt \Nrf$ complex-valued variables and has no specific structure.  In \cite{Zhou2017}, all the factor matrices have a special structure and are determined by only $\Lch$ real-valued variables. In this paper, we derive CRLB in the case where the tensor has  the combination of two structured factor matrices, $\bA$ and $\bC$, and one unstructured factor matrix $\bG$. We also simplify the complicated CRLB expression to a more compact form.

Focusing on the fact that the channel tensor is determined by $\Lch$ AoAs, $\Lch$ path delays, and $\Lch \Tt$ time-varying channel path gains, we define three parameter vectors as  
 $\boldsymbol{\phi} = \begin{bmatrix} \phi_1 & \cdots & \phi_{\Lch} \end{bmatrix}^{\mathsf{T}} $, $\boldsymbol{\tau} = \begin{bmatrix} \tau_1 & \cdots & \tau_{\Lch} \end{bmatrix}^{\mathsf{T}}$, and  $\bg = \mathrm{vec}(\bG) = \begin{bmatrix} g_{1,1} & \cdots & g_{T,\Lch} \end{bmatrix}^{\mathsf{T}} $. 
Let $\boldsymbol{\theta}$  denote a column vector that includes all the parameters such that $\boldsymbol{\theta} = \begin{bmatrix} \boldsymbol{\phi}^{\mathsf{T}} &  \boldsymbol{\tau}^{\mathsf{T}} & \bg^{\mathsf{T}}  & \bg^* \end{bmatrix}^{\mathsf{T}}$. Note that $\bg$ is a complex vector while $\boldsymbol{\phi}$ and $\boldsymbol{\tau}$ are real vectors. 


Since  the analog combining matrix combined with the baseband post-processing matrix is unitary, the elements in the noise tensor $\cN$  become IID circularly symmetric Gaussian with $\cC \cN (0, \sigma^2)$. Consequently, the log-likelihood function of $\boldsymbol{\theta}$ is given by
\begin{equation}
\begin{split}
f(\boldsymbol{\theta} ) 
&= -\Nrf \Kf \Tt  \ln(\pi \sigma^2) - \frac{1}{\sigma^2} \| \bY_{(1)} - \bB (\bG \odot \bC)^{\mathsf{T}} \|^2_{F} \\
&=  -\Nrf \Kf \Tt  \ln(\pi \sigma^2) - \frac{1}{\sigma^2} \| \bY_{(2)} - \bC (\bG \odot \bB)^{\mathsf{T}} \|^2_{F} \\
&= -\Nrf \Kf \Tt  \ln(\pi \sigma^2) - \frac{1}{\sigma^2} \| \bY_{(3)} - \bG (\bC \odot \bB)^{\mathsf{T}} \|^2_{F}.
\end{split}
\end{equation}

Then, the CRLB with respect to the parameter set $\boldsymbol{\theta}$ can be obtained as
\begin{equation}\label{def_CRLB}
\begin{split}
\mathrm{CRLB}(\boldsymbol{\theta})=\boldsymbol{\Omega}^{-1}(\boldsymbol{\theta}), 
\end{split}
\end{equation}
where $\boldsymbol{\Omega}(\boldsymbol{\theta}) \in \mathbb{C}^{2\Lch( \Tt+1) \times 2\Lch( \Tt+1) } $ is the complex Fisher information matrix (FIM) defined as 
\begin{equation}\label{def_FIM}
\begin{split}
\boldsymbol{\Omega}(\boldsymbol{\theta}) &= \mathbb{E} \left[  \frac{\partial  f(\boldsymbol{\theta})}{\partial \boldsymbol{\theta}} \left( \frac{\partial  f(\boldsymbol{\theta})}{\partial \boldsymbol{\theta}} \right)^* \right]. 
\end{split}
\end{equation}
The FIM $\boldsymbol{\Omega}(\boldsymbol{\theta})$ in \eqref{def_FIM} is divided into submatrices as
\begin{equation}\label{def_FIM_r}
\begin{split}
 \boldsymbol{\Omega}(\boldsymbol{\theta})&= \begin{bmatrix} \boldsymbol{\Omega}_{\boldsymbol{\phi} \boldsymbol{\phi} } & \boldsymbol{\Omega}_{\boldsymbol{\phi} \boldsymbol{\tau}} & \boldsymbol{\Omega}_{\boldsymbol{\phi} \bg } & \boldsymbol{\Omega}_{\boldsymbol{\phi} \bg^{\mathsf{C}}} \\
\boldsymbol{\Omega}^*_{\boldsymbol{\phi} \boldsymbol{\tau} } & \boldsymbol{\Omega}_{\boldsymbol{\tau} \boldsymbol{\tau} } & \boldsymbol{\Omega}_{\boldsymbol{\tau} \bg } & \boldsymbol{\Omega}_{\boldsymbol{\tau} \bg^{\mathsf{C}}} \\
\boldsymbol{\Omega}^*_{\boldsymbol{\phi} \bg} & \boldsymbol{\Omega}^*_{\boldsymbol{\tau} \bg } & \boldsymbol{\Omega}_{\bg \bg } & \boldsymbol{\Omega}_{\bg \bg^{\mathsf{C}}} \\
\boldsymbol{\Omega}^*_{\boldsymbol{\phi} \bg^{\mathsf{C}} } & \boldsymbol{\Omega}^*_{\boldsymbol{\tau} \bg^{\mathsf{C}} } & \boldsymbol{\Omega}^*_{\bg \bg^{\mathsf{C}} } & \boldsymbol{\Omega}_{\bg^{\mathsf{C}} \bg^{\mathsf{C}}}
\end{bmatrix}.
\end{split}
\end{equation}
Each submatrix in \eqref{def_FIM_r} is calculated in the following subsections.


\subsection{Calculation of $\boldsymbol{\Omega}_{\boldsymbol{\phi} \boldsymbol{\phi} }  \in \mathbb{C}^{\Lch \times \Lch}$}\label{ssec::phi_phi}

The partial derivative of $f(\boldsymbol{\theta})$ with respect to $\phi_{\ell}$ is given by
\begin{equation}\label{def_phi_phi}
\begin{split}
\frac{\partial  f(\boldsymbol{\theta})}{\partial \phi_{\ell}}
 = \mathrm{Tr}\left( \left( \frac{\partial  f(\boldsymbol{\theta})}{\partial \bB}  \right)^{\mathsf{T}}  \frac{\partial  \bB}{\partial \phi_{\ell}} + \left( \frac{\partial  f(\boldsymbol{\theta})}{\partial \bB^{\mathsf{C}}}  \right)^{\mathsf{T}}  \frac{\partial  \bB^{\mathsf{C}}}{\partial \phi_{\ell}}  \right).
\end{split}
\end{equation}
Let $\acute{\bB} = \begin{bmatrix} \acute{\bb}_{1}(\phi_{1}) & \cdots & \acute{\bb}_{\Lch}(\phi_{\Lch})\end{bmatrix} $ be defined as 
\begin{equation}\label{B_derivative}
\begin{split}
\acute{\bB} 
= \frac{j 2 \pi d_{\mathrm{a}} }{\lambda} & \bW^*   \mathrm{diag}\left(\begin{bmatrix} 0 & 1 & \cdots & \Nt - 1 \end{bmatrix} \right) 
 \begin{bmatrix} \cos(\phi_1) \ba(\phi_1) & \cdots &\cos(\phi_\Lch) \ba(\phi_\Lch) \end{bmatrix},
\end{split} 
\end{equation}
and let $ \bN_{(1)} = \bY_{(1)} - \bB (\bG \odot \bC)^{\mathsf{T}}  $ be the mode-1 unfolding matrix of $\cN$.

By using \eqref{def_phi_phi} and \eqref{B_derivative}, the partial derivative of $f(\boldsymbol{\theta})$ with respect to $\boldsymbol{\phi}$ can be represented as
\begin{equation}\label{def_phi_phi_vec_r1}
\begin{split}
\frac{\partial  f(\boldsymbol{\theta})}{\partial \boldsymbol{\phi}}
 &= \mathrm{diag} \left(  \left( \frac{\partial  f(\boldsymbol{\theta})}{\partial \bB}  \right)^{\mathsf{T}}  \acute{\bB}  +  \left( \frac{\partial  f(\boldsymbol{\theta})}{\partial \bB} \right)^* \acute{\bB} ^{\mathsf{C}} \right) \\
&= \frac{2}{\sigma^2} \mathrm{Re} \left(  \mathrm{diag} \left(  (\bG \odot \bC)^{\mathsf{T}} \bN_{(1)}^* \acute{\bB}\right)  \right) .
\end{split}
\end{equation}
Let $\bV_{\bB} =  \frac{1}{\sigma^2}  (\bG \odot \bC)^{\mathsf{T}} \bN_{(1)}^* \acute{\bB}$ and  $\bd_{\bB}= \mathrm{diag}(\bV_{\bB})$. Then, $\bd_{\bB}$ can be represented as
\begin{equation}\label{def_d_B}
\begin{split}
\bd_{\bB}= \frac{1}{\sigma^2} \left( \acute{\bB} \odot \bG \odot \bC  \right)^{\mathsf{T}} \mathrm{vec}(\bN_{(1)}^*),
\end{split}
\end{equation}
by using the fact that $\mathrm{diag}(\bA^{\mathsf{T}} \bB \bC) = (\bC \odot \bA)^{\mathsf{T}} \mathrm{vec}(\bB)$ when $\bA$ and $\bC$ have the same number of columns. 

Since we consider IID zero mean circularly symmetric complex Gaussian noise, the covariance matrix and the pseudo-covariance matrix of $\mathrm{vec}(\bN_{(1)}^*)$ in \eqref{def_d_B} become
\begin{equation}\label{cov_noise1}
\begin{split}
\mathbb{E} \left[ \mathrm{vec}(\bN_{(1)}^*) \left(\mathrm{vec}(\bN_{(1)}^*)\right)^* \right] 
= \sigma^2 \bI_{\Nrf \Kf \Tt \times \Nrf \Kf \Tt },
\end{split}
\end{equation}
and
\begin{equation}\label{pcov_noise1}
\begin{split}
 \mathbb{E} \left[ \mathrm{vec}(\bN_{(1)}^*) \left(\mathrm{vec}(\bN_{(1)}^*)\right)^{\mathsf{T}} \right] = \mathbf{0}_{\Nrf \Kf \Tt \times \Nrf \Kf \Tt }.
\end{split}
\end{equation}

From \eqref{cov_noise1} and \eqref{pcov_noise1}, the covariance matrix of the complex vector $\bd_{\bB}$ becomes
\begin{equation}\label{cov_v_B}
\begin{split}
\bC_{\bd_{\bB}} &= \mathbb{E} \left[ \bd_{\bB} \bd_{\bB}^* \right] \\
&=  \frac{1}{\sigma^2} \left( \acute{\bB} \odot \bG \odot \bC  \right)^{\mathsf{T}}   \left(\acute{\bB} \odot \bG \odot \bC   \right)^{\mathsf{C}} \\
&=  \frac{1}{\sigma^2} \left(  \acute{\bB}^*  \acute{\bB}   \circledcirc  \bG^*\bG \circledcirc \bC^* \bC   \right)^{\mathsf{T}},
\end{split}
\end{equation}
and the pseudo-covariance matrix  becomes $\tilde{\bC}_{\bd_{\bB}} = \mathbb{E} \left[ \bd_{\bB} \bd_{\bB}^{\mathsf{T}} \right] = \mathbf{0} $.
Consequently, the submatrix $ \boldsymbol{\Omega}_{\boldsymbol{\phi} \boldsymbol{\phi} } $ becomes
\begin{equation} \label{def_Omega_phi_phi_vec}
\begin{split}
\boldsymbol{\Omega}_{\boldsymbol{\phi} \boldsymbol{\phi} } 
&= \mathbb{E} \left[ \frac{\partial  f(\boldsymbol{\theta})}{\partial \boldsymbol{\phi}} \left( \frac{\partial  f(\boldsymbol{\theta})}{\partial \boldsymbol{\phi}}  \right)^* \right] \\
&= \mathbb{E} \left[\left( \bd_{\bB} + \bd_{\bB}^{\mathsf{C}} \right)   ( \bd_{\bB}^* + \bd_{\bB}^{\mathsf{T}} ) \right] \\
&= 2\mathrm{Re} \left(\bC_{\bd_{\bB}} \right)  + 2\mathrm{Re} \left(\tilde{\bC}_{\bd_{\bB}}  \right)  \\
&= \frac{2}{\sigma^2} \mathrm{Re} \left( \left(  \acute{\bB}^*  \acute{\bB}   \circledcirc  \bG^*\bG \circledcirc \bC^* \bC   \right)^{\mathsf{T}} \right)  .
\end{split}
\end{equation}

\subsection{Calculation of $\boldsymbol{\Omega}_{\boldsymbol{\tau} \boldsymbol{\tau} }  \in \mathbb{C}^{\Lch \times \Lch}$}\label{ssec::tau_tau}

Let $\acute{\bc}_{\ell}(\tau_{\ell}) \in \mathbb{C}^{\Kf \times 1}$ be a vector whose $k$-th element is defined as 
\begin{equation}\label{c_derivative}
\begin{split}
[ \acute{\bc}_{\ell}(\tau_{\ell})]_k = - \sum_{d=0}^{N_{\mathrm{CP}}-1} p^{'}_{\mathrm{PS}}(dT_s - \tau_{\ell}) e^{ -\frac{j 2 \pi (k-1) d}{\Kf}},
\end{split}
\end{equation}
where $p^{'}_{\mathrm{PS}}(x)$ is the first derivative of $p_{\mathrm{PS}}(x)$.  
Then, the partial derivative of $f(\boldsymbol{\theta})$ with respect to $\boldsymbol{\tau}$ can be obtained as 
\begin{equation}\label{f_tau}
\begin{split}
\frac{\partial  f(\boldsymbol{\theta})}{\partial \boldsymbol{\tau}} 
&= \frac{2}{\sigma^2} \mathrm{Re} \left( \mathrm{diag}\left( (\bG \odot \bB)^{\mathsf{T}} \left(  \bY_{(2)}^* - (\bG \odot \bB)^{\mathsf{C}} \bC^*   \right) \acute{\bC} \right)  \right),
\end{split}
\end{equation}
where $\acute{\bC} = \begin{bmatrix} \acute{\bc}_{1}(\tau_{1}) & \cdots & \acute{\bc}_{\Lch}(\tau_{\Lch})\end{bmatrix} $.
Let $ \bN_{(2)} = \bY_{(2)} - \bC (\bG \odot \bB)^{\mathsf{T}}  $ be the mode-2 unfolding matrix of $\cN$, and $\bV_{\bC} =  \frac{1}{\sigma^2}  (\bG \odot \bB)^{\mathsf{T}} \bN_{(2)}^* \acute{\bC}$.
Let $\bd_{\bC}$ denote  
\begin{equation}\label{def_d_C}
\begin{split}
\bd_{\bC} &= \mathrm{diag}(\bV_{\bC}) = \frac{1}{\sigma^2} \left( \acute{\bC} \odot \bG \odot \bB  \right)^{\mathsf{T}} \mathrm{vec}(\bN_{(2)}^*).
\end{split}
\end{equation}

Similarly to \sref{ssec::phi_phi},  the submatrix $ \boldsymbol{\Omega}_{\boldsymbol{\tau} \boldsymbol{\tau} }$ can be calculated as 
\begin{equation} \label{def_Omega_tau_tau_vec}
\begin{split}
\boldsymbol{\Omega}_{\boldsymbol{\tau} \boldsymbol{\tau} } 
&= \frac{2}{\sigma^2} \mathrm{Re} \left( \left(  \acute{\bC}^*  \acute{\bC}   \circledcirc  \bG^*\bG \circledcirc \bB^* \bB   \right)^{\mathsf{T}} \right) . 
\end{split}
\end{equation}

\subsection{Calculation of $\boldsymbol{\Omega}_{\bg \bg } \in \mathbb{C}^{\Tt \Lch \times \Tt \Lch}$, $\boldsymbol{\Omega}_{\bg \bg^{\mathsf{C}} }\in \mathbb{C}^{\Tt \Lch \times \Tt \Lch}$, and $\boldsymbol{\Omega}_{\bg^{\mathsf{C}} \bg^{\mathsf{C}} }\in \mathbb{C}^{\Tt \Lch \times \Tt \Lch}$}\label{ssec::g_g}


From the fact that $\mathbb{E} \left[ \mathrm{vec}(\bN_{(3)}^{\mathsf{C}}) \left(\mathrm{vec}(\bN_{(3)}^{\mathsf{C}})\right)^* \right] 
= \sigma^2 \bI_{\Nrf \Kf \Tt \times \Nrf \Kf \Tt } $, the submatrix $\boldsymbol{\Omega}_{\bg \bg } $  becomes
\begin{equation} \label{g_g}
\begin{split}
\boldsymbol{\Omega}_{\bg \bg }
&= \mathbb{E} \left[ \frac{\partial  f(\boldsymbol{\theta})}{\partial \mathrm{vec}( \bG )} \left( \frac{\partial  f(\boldsymbol{\theta})}{\partial  \mathrm{vec}( \bG )}  \right)^* \right] \\
&=  \frac{1}{\sigma^4} \left(   (\bC \odot \bB)^{\mathsf{T}} \otimes \bI_\Tt \right)  \mathbb{E} \left[ \mathrm{vec}(\bN_{(3)}^{\mathsf{C}}) \left(\mathrm{vec}(\bN_{(3)}^{\mathsf{C}})\right)^* \right]   \left(   (\bC \odot \bB)^{\mathsf{T}} \otimes \bI_\Tt \right)^* \\
&=  \frac{1}{\sigma^2} \left( \bC^* \bC \circledcirc  \bB^* \bB \right)^{\mathsf{T}} \otimes \bI_\Tt  .
\end{split}
\end{equation}

Since $ \mathbb{E} \left[ \mathrm{vec}(\bN_{(3)}^{\mathsf{C}}) \left(\mathrm{vec}(\bN_{(3)}^{\mathsf{C}})\right)^{\mathsf{T}} \right] 
= \mathbf{0}_{\Nrf \Kf \Tt \times \Nrf \Kf \Tt } $, the submatrix $\boldsymbol{\Omega}_{\bg \bg^{\mathsf{C}} } $ is given by 
\begin{equation} \label{g_g_C}
\begin{split}
\boldsymbol{\Omega}_{\bg \bg^{\mathsf{C}} }
&= \mathbb{E} \left[ \frac{\partial  f(\boldsymbol{\theta})}{\partial \mathrm{vec}( \bG )} \left( \frac{\partial  f(\boldsymbol{\theta})}{\partial  \mathrm{vec}( \bG^{\mathsf{C}} )}  \right)^* \right] \\
&=  \frac{1}{\sigma^4} \left(   (\bC \odot \bB)^{\mathsf{T}} \otimes \bI_\Tt \right) \mathbb{E} \left[ \mathrm{vec}(\bN_{(3)}^{\mathsf{C}}) \left(\mathrm{vec}(\bN_{(3)}^{\mathsf{C}})\right)^{\mathsf{T}} \right]   \left(   (\bC \odot \bB)^{\mathsf{T}} \otimes \bI_\Tt \right)^{\mathsf{T}} \\
&=  \mathbf{0}_{\Tt \Lch \times \Tt \Lch} .
\end{split}
\end{equation}

The submatrix $\boldsymbol{\Omega}_{\bg^{\mathsf{C}} \bg^{\mathsf{C}} }$ can be obtained from $\boldsymbol{\Omega}_{\bg \bg }$   such that $\boldsymbol{\Omega}_{\bg^{\mathsf{C}} \bg^{\mathsf{C}} } = \boldsymbol{\Omega}_{\bg \bg }^{\mathsf{C}}$.


\subsection{Calculation of $\boldsymbol{\Omega}_{\boldsymbol{\phi} \boldsymbol{\tau} } \in \mathbb{C}^{\Lch \times \Lch}$}\label{ssec::phi_tau}

The submatrix $ \boldsymbol{\Omega}_{\boldsymbol{\phi} \boldsymbol{\tau} }$ is given by
\begin{equation}\label{phi_tau_0}
\begin{split}
\boldsymbol{\Omega}_{\boldsymbol{\phi} \boldsymbol{\tau} } 
&= \mathbb{E} \left[ \frac{\partial  f(\boldsymbol{\theta})}{\partial \boldsymbol{ \phi}} \left( \frac{\partial  f(\boldsymbol{\theta})}{\partial  \boldsymbol{\tau}}  \right)^* \right] \\
&= \mathbb{E} \left[  \left(\bd_{\bB} + \bd_{\bB}^{\mathsf{C}} \right)  \left( \bd_{\bC} + \bd_{\bC}^{\mathsf{C}} \right)^* \right] \\
&= 2\mathrm{Re} \left(\bC_{\bd_{\bB}, \bd_{\bC}} \right) + 2\mathrm{Re} \left(\tilde{\bC}_{\bd_{\bB}, \bd_{\bC}}  \right),
\end{split}
\end{equation}
where $\bC_{\bd_{\bB}, \bd_{\bC}} = \mathbb{E} \left[ \bd_{\bB} \bd_{\bC}^* \right]$ and $\tilde{\bC}_{\bd_{\bB}, \bd_{\bC}} = \mathbb{E} \left[ \bd_{\bB} \bd_{\bC}^{\mathsf{T}} \right]$.

To calculate $\bC_{\bd_{\bB}, \bd_{\bC}}$, let us first start with calculating the cross-covariance matrix of $ \mathrm{vec}(\bN_{(1)}^*)$ and $ \mathrm{vec}(\bN_{(2)}^*)$, which are associated with  the mode-1 and mode-2 unfolding matrix of $\cN$. 
Let  $\be_i \in \mathbb{C}^{\Nrf \Kf \Tt \times 1}$ be the $i$-th unit coordinate vector and  $\bC_{\bn_{(1)},\bn_{(2)}} =\mathbb{E} \left[  \mathrm{vec}(\bN_{(1)}^*)  \left( \mathrm{vec}(\bN_{(2)}^*) \right)^* \right]$.
Using the fact that $[\cN]_{m,k,t}$ is expressed in different ways  as
\begin{equation}\label{N_indexing}
\begin{split}
[\cN]_{m,k,t} 
&= [\mathrm{vec}\left( \bN_{(1)} \right)]_{ k+(t-1)\Kf + (m-1) \Kf \Tt } \\
&= [\mathrm{vec}\left( \bN_{(2)} \right)]_{ m+(t-1)\Nrf + (k-1) \Nrf \Tt } ,
\end{split}
\end{equation}
the cross-covariance matrix $\bC_{\bn_{(1)},\bn_{(2)}} $ can be represented as 
\begin{equation}\label{cov_n1_n2}
\begin{split}
\bC_{\bn_{(1)},\bn_{(2)}} &=\mathbb{E} \left[  \mathrm{vec}(\bN_{(1)}^*)  \left( \mathrm{vec}(\bN_{(2)}^*) \right)^* \right] \\
&= \sigma^2 \sum_{m=1}^\Nrf  \sum_{k=1}^\Kf  \sum_{t=1}^\Tt  \be_{ k+(t-1)\Kf + (m-1)\Kf \Tt } \be^{\mathsf{T}}_{ m+(t-1)\Nrf + (k-1)\Nrf \Tt  }.
\end{split}
\end{equation}
 Consequently, $\bC_{\bn_{(1)},\bn_{(2)}} \in \mathbb{C}^{\Nrf \Kf \Tt  \times \Nrf \Kf \Tt }$ is a matrix that has only $\Nrf \Kf \Tt $ nonzero elements whose amplitudes are equal to $\sigma^2$. 
From \eqref{cov_n1_n2},  the cross-covariance matrix of $\bd_{\bB}$ and $\bd_{\bC}$ can be expressed as
\begin{equation}\label{cov_d_B_d_C_H}
\begin{split}
\bC_{\bd_{\bB}, \bd_{\bC}} 
&=  \frac{1}{\sigma^4} \left( \acute{\bB} \odot \bG \odot \bC  \right)^{\mathsf{T}} \bC_{\bn_{(1)}, \bn_{(2)}}   \left( \acute{\bC} \odot \bG \odot \bB  \right)^{\mathsf{C}} \\
&= \frac{1}{\sigma^2}\left(\bB^* \acute{\bB}  \circledcirc \bG^* \bG \circledcirc   \acute{\bC}^* \bC \right)^{\mathsf{T}}.
\end{split}
\end{equation}
Since $\tilde{\bC}_{\bn_{(1)},\bn_{(2)}} =\mathbb{E} \left[  \mathrm{vec}(\bN_{(1)}^*)  \left( \mathrm{vec}(\bN_{(2)}^*) \right)^{\mathsf{T}} \right] =\mathbf{0}$, the pseudo-cross-covariance matrix of $\bd_{\bB}$ and $\bd_{\bC}$ becomes 
\begin{equation}\label{cov_d_B_d_C_T}
\begin{split}
\tilde{\bC}_{\bd_{\bB}, \bd_{\bC}} &= \mathbb{E} \left[ \bd_{\bB} \bd_{\bC}^{\mathsf{T}} \right] = \mathbf{0}.
\end{split}
\end{equation}

From \eqref{cov_d_B_d_C_H} and \eqref{cov_d_B_d_C_T}, the submatrix $ \boldsymbol{\Omega}_{\boldsymbol{\phi} \boldsymbol{\tau} }$   in \eqref{phi_tau_0} can be rewritten as
\begin{equation}
\begin{split}
 \boldsymbol{\Omega}_{\boldsymbol{\phi} \boldsymbol{\tau} }  
&=  \frac{2}{\sigma^2} \mathrm{Re} \left( \left(\bB^* \acute{\bB}  \circledcirc \bG^* \bG \circledcirc   \acute{\bC}^* \bC \right)^{\mathsf{T}}  \right). 
\end{split}
\end{equation}

\subsection{Calculation of $\boldsymbol{\Omega}_{\boldsymbol{\phi} \bg } \in \mathbb{C}^{\Lch \times \Tt \Lch}$ and $\boldsymbol{\Omega}_{\boldsymbol{\phi} \bg^{\mathsf{C}} } \in \mathbb{C}^{\Lch \times \Tt \Lch}$}\label{ssec::phi_g}

 The submatrix $ \boldsymbol{\Omega}_{\boldsymbol{\phi} \bg }  $ is expressed as
\begin{equation}\label{phi_g}
\begin{split}
 \boldsymbol{\Omega}_{\boldsymbol{\phi} \bg } 
&= \mathbb{E} \left[ \frac{\partial  f(\boldsymbol{\theta})}{\partial \boldsymbol{\phi}} \left(   \frac{\partial  f(\boldsymbol{\theta})}{\partial  \mathrm{vec}( \bG )}   \right)^*\right] \\
&=\frac{1}{\sigma^2} \mathbb{E} \left[  \left( \bd_{\bB} + \bd_{\bB}^{\mathsf{C}} \right)    \left(\mathrm{vec}(\bN_{(3)}^{\mathsf{C}})\right)^*  \left(   (\bC \odot \bB)^{\mathsf{T}} \otimes \bI_\Tt \right)^*   \right] \\
&=\frac{1}{\sigma^4} \left(  \left( \acute{\bB} \odot \bG \odot \bC  \right)^{\mathsf{T}} \bC_{\bn_{(1)}, \bn^{\mathsf{C}}_{(3)}} +  \left( \acute{\bB} \odot \bG \odot \bC  \right)^* \tilde{\bC}_{\bn_{(1)}, \bn^{\mathsf{C}}_{(3)}} \right) \cdot \left(   (\bC \odot \bB)^{\mathsf{C}} \otimes \bI_\Tt \right),  \\
\end{split}
\end{equation}
where  $\tilde{\bC}_{\bn_{(1)},\bn^{\mathsf{C}}_{(3)}} =\mathbb{E} \left[  \mathrm{vec}(\bN_{(1)})  \left( \mathrm{vec}(\bN_{(3)}) \right)^{\mathsf{T}} \right] = \mathbf{0}$  and
\begin{equation}\label{C_n1_n3_T}
\begin{split}
\bC_{\bn_{(1)},\bn^{\mathsf{C}}_{(3)}}  &=\mathbb{E} \left[  \mathrm{vec}(\bN_{(1)}^*)  \left( \mathrm{vec}(\bN_{(3)}) \right)^{\mathsf{T}} \right] \\
&= \sigma^2 \sum_{m=1}^\Nrf  \sum_{k=1}^\Kf  \sum_{t=1}^\Tt  \be_{ k+(t-1)\Kf + (m-1)\Kf \Tt }  \be^{\mathsf{T}}_{ t+ (m+(k-1)\Nrf -1)\Tt }.
\end{split}
\end{equation}

By using \eqref{C_n1_n3_T},  we can further simplify $ \boldsymbol{\Omega}_{\boldsymbol{\phi} \bg } $ in  \eqref{phi_g}  as
\begin{equation}\label{phi_g_r1}
\begin{split}
 \boldsymbol{\Omega}_{\boldsymbol{\phi} \bg }
&=\frac{1}{\sigma^4}  \left( \acute{\bB} \odot \bG \odot \bC  \right)^{\mathsf{T}} \bC_{\bn_{(1)}, \bn^{\mathsf{C}}_{(3)}}  \left(   (\bC \odot \bB)^{\mathsf{C}} \otimes \bI_\Tt \right)  \\
&=\frac{1}{\sigma^2}   \sum_{t=1}^\Tt  \left(\left( \left(\bB^* \acute{\bB}\right) \circledcirc  \left( \bC^* \bC \right) \circledcirc  \left( \mathbf{1}_\Lch  [\bG]_{t,:} \right) \right)  \otimes \be_t \right)^{\mathsf{T}}  \\
&=\frac{1}{\sigma^2} \left( \left(\bB^* \acute{\bB}\circledcirc \bC^* \bC \right) \odot \bG \right)^{\mathsf{T}}.
\end{split}
\end{equation}

The submatrix $ \boldsymbol{\Omega}_{\boldsymbol{\phi} \bg^{\mathsf{C}} } $ is expressed as $\boldsymbol{\Omega}_{\boldsymbol{\phi} \bg^{\mathsf{C}} }=   \boldsymbol{\Omega}^{\mathsf{C}}_{\boldsymbol{\phi} \bg } $.

\subsection{Calculation of $\boldsymbol{\Omega}_{\boldsymbol{\tau} \bg } \in \mathbb{C}^{\Lch \times \Tt \Lch}$ and $\boldsymbol{\Omega}_{\boldsymbol{\tau} \bg^{\mathsf{C}} } \in \mathbb{C}^{\Lch \times \Tt \Lch}$}\label{ssec::tau_g}

Similar to \sref{ssec::phi_g}, we can obtain 
$ \boldsymbol{\Omega}_{\boldsymbol{\tau} \bg } $  and $\boldsymbol{\Omega}_{\boldsymbol{\tau} \bg^{\mathsf{C}} }$ as 
\begin{equation}\label{tau_g}
\begin{split}
 \boldsymbol{\Omega}_{\boldsymbol{\tau} \bg } 
&= \mathbb{E} \left[ \frac{\partial  f(\boldsymbol{\theta})}{\partial \boldsymbol{\tau}} \left(  \frac{\partial  f(\boldsymbol{\theta})}{\partial  \mathrm{vec}( \bG )}   \right)^*\right] \\
&=\frac{1}{\sigma^2} \left( \left(\bC^* \acute{\bC}\circledcirc \bB^* \bB \right) \odot \bG \right)^{\mathsf{T}},
\end{split}
\end{equation}
and  $\boldsymbol{\Omega}_{\boldsymbol{\tau} \bg^{\mathsf{C}} }=   \boldsymbol{\Omega}^{\mathsf{C}}_{\boldsymbol{\tau} \bg } $.

\subsection{CRLB for the $\phi$ estimation}\label{ssec::CRLB_phi}

The results of the preceding subsections are summarized as
\begin{equation}\label{summary_submatrices}
\begin{split}
\boldsymbol{\Omega}_{\boldsymbol{\phi} \boldsymbol{\phi} } 
&= \frac{2}{\sigma^2} \mathrm{Re} \left( \left(  \acute{\bB}^*  \acute{\bB}   \circledcirc \bC^* \bC  \circledcirc  \bG^*\bG  \right)^{\mathsf{T}} \right),  \\
\boldsymbol{\Omega}_{\boldsymbol{\tau} \boldsymbol{\tau} } 
&= \frac{2}{\sigma^2} \mathrm{Re} \left(\left(  \bB^* \bB  \circledcirc   \acute{\bC}^*  \acute{\bC}   \circledcirc  \bG^*\bG \right)^{\mathsf{T}} \right)  ,\\
\boldsymbol{\Omega}_{\boldsymbol{\phi} \boldsymbol{\tau} } 
&= \frac{2}{\sigma^2} \mathrm{Re} \left(\left(  \bB^* \acute{\bB}  \circledcirc   \acute{\bC}^*  \bC  \circledcirc  \bG^*\bG \right)^{\mathsf{T}} \right) , \\
\boldsymbol{\Omega}_{\bg \bg } &=  \frac{1}{\sigma^2} \left(  \bB^* \bB  \circledcirc  \bC^* \bC \right)^{\mathsf{T}} \otimes \bI_\Tt , \\
\boldsymbol{\Omega}_{\boldsymbol{\phi} \bg } &=\frac{1}{\sigma^2} \left( \left(\bB^* \acute{\bB}\circledcirc \bC^* \bC \right) \odot \bG \right)^{\mathsf{T}} ,\\
\boldsymbol{\Omega}_{\boldsymbol{\tau} \bg } &=\frac{1}{\sigma^2} \left( \left(\bB^* \bB \circledcirc \bC^* \acute{\bC} \right) \odot \bG \right)^{\mathsf{T}} .
\end{split}
\end{equation}



Letting $\boldsymbol{\Omega}_1 = \begin{bmatrix}  \boldsymbol{\Omega}_{\boldsymbol{\phi} \boldsymbol{\tau}} & \boldsymbol{\Omega}_{\boldsymbol{\phi} \bg } & \boldsymbol{\Omega}_{\boldsymbol{\phi} \bg} ^{\mathsf{C}}
\end{bmatrix}$ and
\begin{equation}\label{def_Omega2}
\begin{split}
\boldsymbol{\Omega}_2 &= \begin{bmatrix} \boldsymbol{\Omega}_{\boldsymbol{\tau} \boldsymbol{\tau} } & \boldsymbol{\Omega}_{\boldsymbol{\tau} \bg } & \boldsymbol{\Omega}_{\boldsymbol{\tau} \bg}^{\mathsf{C}} \\
 \boldsymbol{\Omega}^*_{\boldsymbol{\tau} \bg } & \boldsymbol{\Omega}_{\bg \bg } & \mathbf{0} \\
 \boldsymbol{\Omega}^{\mathsf{T}}_{\boldsymbol{\tau} \bg } & \mathbf{0} & \boldsymbol{\Omega}_{\bg \bg}^{\mathsf{C}} 
\end{bmatrix},
\end{split}
\end{equation}
 the CRLB for the $\phi_{\ell}$ estimation can be expressed in a compact form as
\begin{equation}\label{def_CRLB_phi}
\begin{split}
\mathrm{CRLB}(\phi_{\ell}) = \left[ \left( \boldsymbol{\Omega}_{\boldsymbol{\phi} \boldsymbol{\phi} } - \boldsymbol{\Omega}_1^* \boldsymbol{\Omega}_2^{-1} \boldsymbol{\Omega}_1 \right)^{-1} \right]_{\ell,\ell},
\end{split}
\end{equation}
by using the Schur complement and the matrix inversion lemma.

\section{Comparison with CS-based or MUSIC-based methods}\label{sec:Comparison_others}

In this section, we explain  two other approaches that estimate spatial channel covariance or subspace for comparison:  
1) CS-based methods and 2) MUSIC-based methods. 

\subsection{Prior work based on CS}\label{ssec::prior_CS_based}

The channel frequency response vector in \eqref{WB_ch_model_CFR} can be represented by using a matrix form as
\begin{equation}\label{WB_ch_model_CFR_matrixForm}
\begin{split}
\bh_{t,k}  
&= \bA  \left(\mathring{\bg}_t  \circledcirc \mathring{\bc}_k \right)
,
\end{split}
\end{equation}
where $\mathring{\bg}_t = [\bG^{\mathsf{T}}]_{:,t} \in \mathbb{C}^{\Lch \times 1}$ and $\mathring{\bc}_k = [\bC^{\mathsf{T}}]_{:,k} \in \mathbb{C}^{\Lch \times 1}$.
Let $\bA_{\mathrm{D}} \in \mathbb{C}^{\Nt \times \Dg}$ be a dictionary matrix whose $\Dg$ columns are composed of the array response vectors associated with a predefined set of AoAs. 
In the CS framework, the channel model in \eqref{WB_ch_model_CFR_matrixForm} is rewritten as
\begin{equation}\label{WB_ch_model_CFR_matrixForm2}
\begin{split}
\bh_{t,k}  \approx \bA_{\mathrm{D}}  \left(\mathring{\bg}_{\mathrm{D},t}  \circledcirc \mathring{\bc}_{\mathrm{D},k} \right),
\end{split}
\end{equation}
where  $\mathring{\bg}_{\mathrm{D},t} \in \mathbb{C}^{\Dg \times 1}$ and $\mathring{\bc}_{\mathrm{D},k}  \in \mathbb{C}^{\Dg \times 1}$ are sparse column vectors with $\Lch$ nonzero elements of $\mathring{\bg}_t$ and $\mathring{\bc}_k$ in the space domain.
The positions of the $\Lch$ nonzero elements indicate  AoAs, and thus $\mathring{\bg}_{\mathrm{D},t}$ and $\mathring{\bc}_{\mathrm{D},k} $ share the same support for all $t$ and $k$. To exploit the joint sparsity of $\mathring{\bg}_{\mathrm{D},t}$ and $\mathring{\bc}_{\mathrm{D},k} $, we can use CS techniques known as multiple measurement vector (MMV) problems instead of conventional single measurement vector (SMV) problems \cite{CotterRaoEnganDelgado2005::6, ChenTSP2006,Determe2016}. While simultaneous orthogonal matching pursuit (SOMP) is known as an adequate algorithm for the general MMV problems, a more advanced CS algorithm was proposed for the spatial channel covariance estimation problem in \cite{Park2017twc_Cov}. We will compare the algorithm in \cite{Park2017twc_Cov} to our tensor-based method.

\subsection{Prior work based on MUSIC}\label{ssec::prior_MUSIC_based}

In conventional fully-digital architectures, the goal of the MUSIC algorithm is to estimate AoAs from the spatial channel covariance matrix. In other words, the covariance must be known prior to applying the MUSIC algorithms. Note that the spatial channel covariance can be estimated from the covariance of the received signal vectors in fully-digital architectures. 
Although the spatial channel covariance estimation is not straightforward in the hybrid architectures, the MUSIC algorithm can be applied to the subspace estimation problem for hybrid architectures. 
It is worthwhile to note that only the subspace can be estimated and the covariance cannot be estimated by using the MUSIC-based approach. Since the subspace is enough for the hybrid precoder design in some cases as in SU-MIMO systems, we will compare our proposed work with the MUSIC-based method in terms of subspace estimation.  
The overall process for the MUSIC-based method is composed of three steps. First, the sample covariance of the baseband received signal vectors $\by_{t,k}$ in \eqref{rx_signal_vector_tk} is estimated for all $t$ and $k$ as 
\begin{equation}\label{cov_bb_rx}
\begin{split}
\bR_{\by}   &= \frac{1}{\Tt \Kf} \sum_{t=1}^\Tt \sum_{k=1}^\Kf \by_{t,k} \by_{t,k}^* \\
&= \frac{1}{\Tt \Kf} \bW^* \bA \left( \bG^* \bG  \circledcirc  \bC^* \bC \right) \bA^* \bW +  \bR_{\bn},
\end{split}
\end{equation}
where $\bR_{\bn} = \frac{1}{\Tt \Kf}  \left( \sum_{t=1}^\Tt \sum_{k=1}^\Kf \bn_{t,k} \bn_{t,k}^* \right)$.
Let the SVD of $\bR_{\by}  $ be 
\begin{equation}\label{cov_bb_rx_SVD}
\begin{split}
\bR_{\by}   &= \bU_{\bx}  \mathbf{\Sigma}_{\bx}   \bU_{\bx}^* +  \bU_{\bn}  \mathbf{\Sigma}_{\bn}   \bU_{\bn}^*,
\end{split}
\end{equation}
where $\bU_{\bx} \in \mathbb{C}^{\Nrf \times \Lch}$ is the signal subspace and $\bU_{\bn} \in \mathbb{C}^{\Nrf \times (\Nrf - \Lch)}$ is the subspace orthogonal to the signal subspace. Let $\bb_{\bW}(\phi) = \bW^* \ba(\phi)$. Since the RF chains can be regarded as the effective antennas from the viewpoint of the estimator at baseband, the vector $\bb_{\bW}(\phi)$ can be considered as  the {\it{effective}} array response vector with a reduced size.   
The second step is to find the $\Lch$ highest peaks of the function of $\phi$ defined  as
\begin{equation}\label{MUSIC_function}
\begin{split}
f_{{\mathrm{MUSIC}}} (\phi) = \frac{1}{ \bb_{\bW}^*(\phi)\bU_{\bn} \bU_{\bn}^* \bb_ {\bW}(\phi)}. 
\end{split}
\end{equation}
The final step is to reconstruct the subspace of the channel by using the $\phi_{\ell}$s for $\ell =1,...,\Lch$ that are obtained from the subspace of the baseband received signals. The subspace of the channel is given by the subspace of $\begin{bmatrix} \ba(\phi_{1}) & \cdots \ba(\phi_{\Lch}) \end{bmatrix}$.

The CRLB for $\phi_{\ell}$s in the MUSIC-based method case is given by \cite{StoicaNehorai1989}
\begin{equation}\label{MUSIC_CRLB}
\begin{split}
\mathrm{CRLB}_{\mathrm{MUSIC}}(\boldsymbol{\phi}) 
&= 
\frac{\sigma}{2} \left( \sum_{t=1}^\Tt \sum_{k=1}^\Kf \mathrm{Re} \left( \bZ_{t,k}^* \acute{\bB}^* \left(\bI - \bB \left( \bB^* \bB \right)^{-1} \bB^* \right) \acute{\bB} \bZ_{t,k} \right) \right)  ^{ -1} , 
\end{split}
\end{equation}
where $\acute{\bB}$ is defined in \eqref{B_derivative} and $\bZ_{t,k}$ is defined as $\bZ_{t,k} = \mathrm{diag} \left(  [\bG^{\mathsf{T}}]_{:,t} \circledcirc  [\bC^{\mathsf{T}}]_{:,k} \right)$. The  bound of the RPE in the MUSIC-based method case can also be obtained from \eqref{eff_metric_mean}  as in the tensor-based method case.

\begin{figure}[!t]
	\centering
	\subfigure[center][{MSE of $\phi$}]{
		\includegraphics[width=0.44\columnwidth]{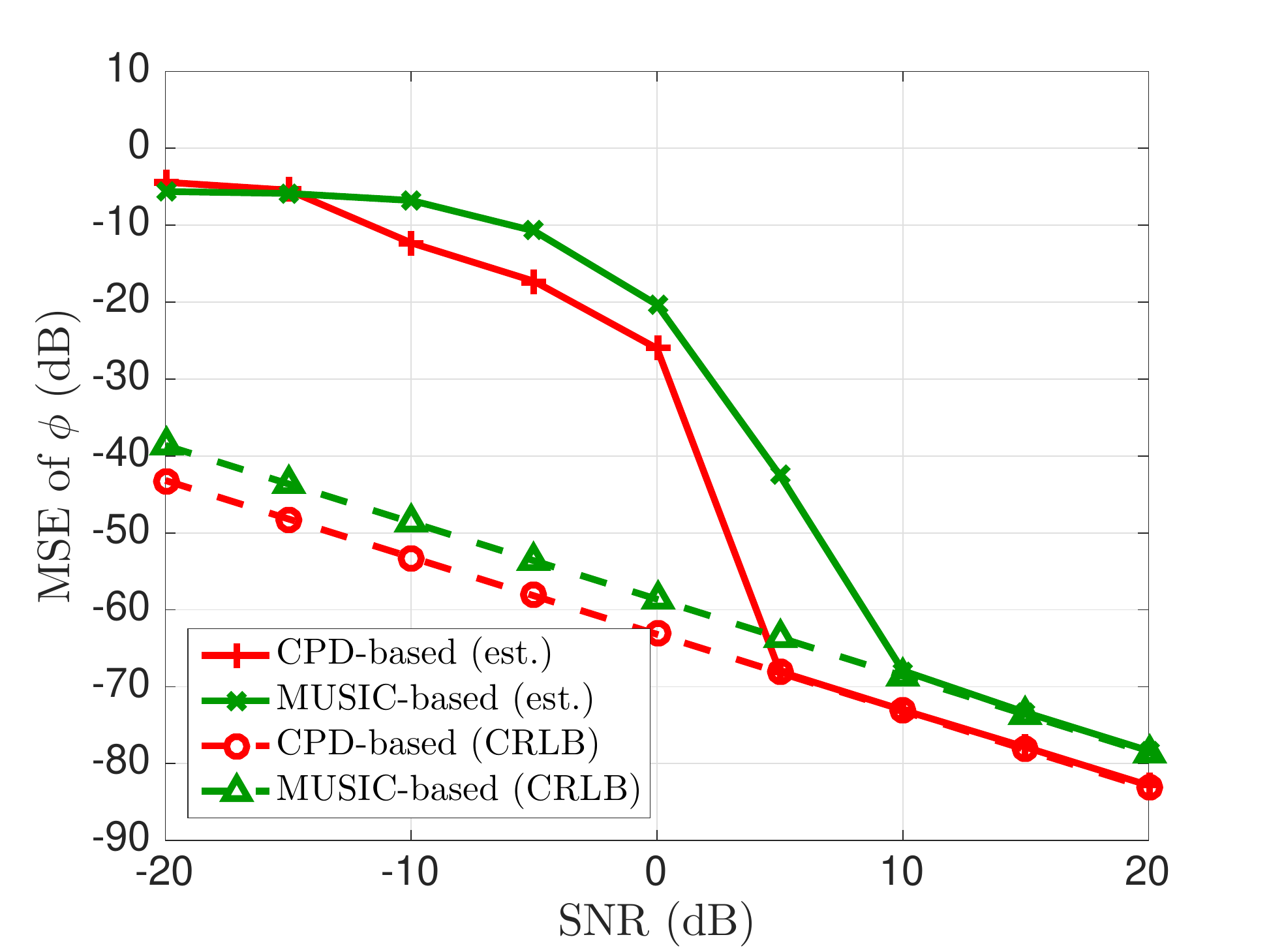}
		\label{fig:CRLB_phi_wMUSIC}}
	\subfigure[center][{1-$\mathbb{E}[\eta]$}]{
		\includegraphics[width=0.44\columnwidth]{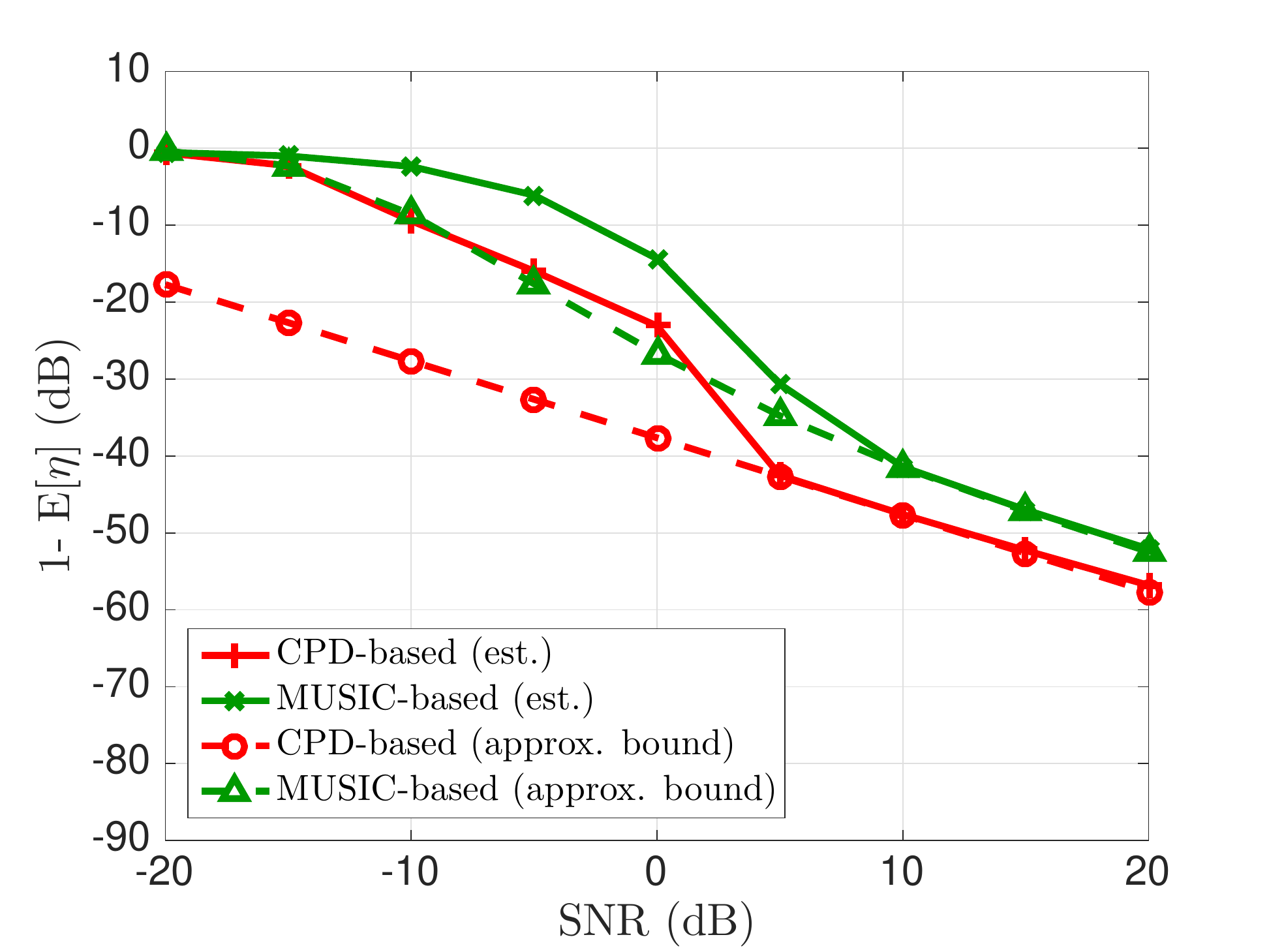}
		\label{fig:CRLB_EffMetric_wMUSIC}}
	\caption{  Comparison between the proposed CPD-based method and the MUSIC-based method in terms of the MSE of $\phi$ and 1-$\mathbb{E}[\eta]$  when $\Tt=20$, $ \Kf=128$, $\Nt=64$, $\Nrf=8$, and $\Lch=6$. }
	\label{fig:comparison_MUSIC_performance_metric}
\end{figure}

\section{Simulation results}\label{sec:simulation_results}

In this section, we numerically evaluate the CRLB analysis in \sref{sec:CRLB}. We also present simulation results 
to   demonstrate the performance of the proposed spatial channel covariance
estimation algorithms based on CPD of higher-order tensors.


\subsection{Analytical results on CRLB}\label{ssec::CRLB_sim}

\begin{figure}[!t]
	\centering
	\subfigure[center][{CRLB($\phi$) vs. $\Tt$}]{
		\includegraphics[width=0.44\columnwidth]{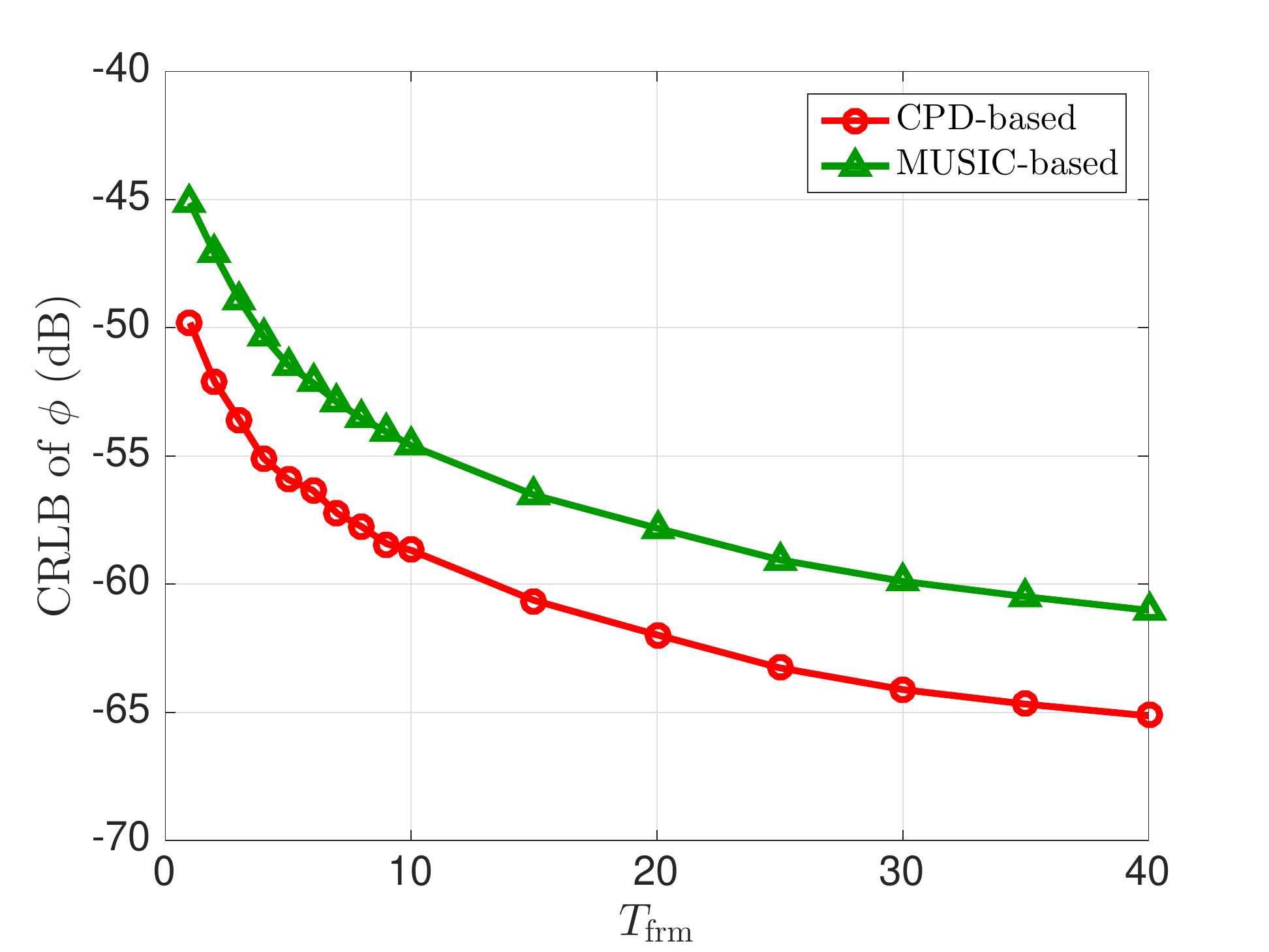}
		\label{fig:CRLB_phi_fixed_vsT}}
	\subfigure[center][{CRLB($\phi$) vs. $\Kf$}]{
		\includegraphics[width=0.44\columnwidth]{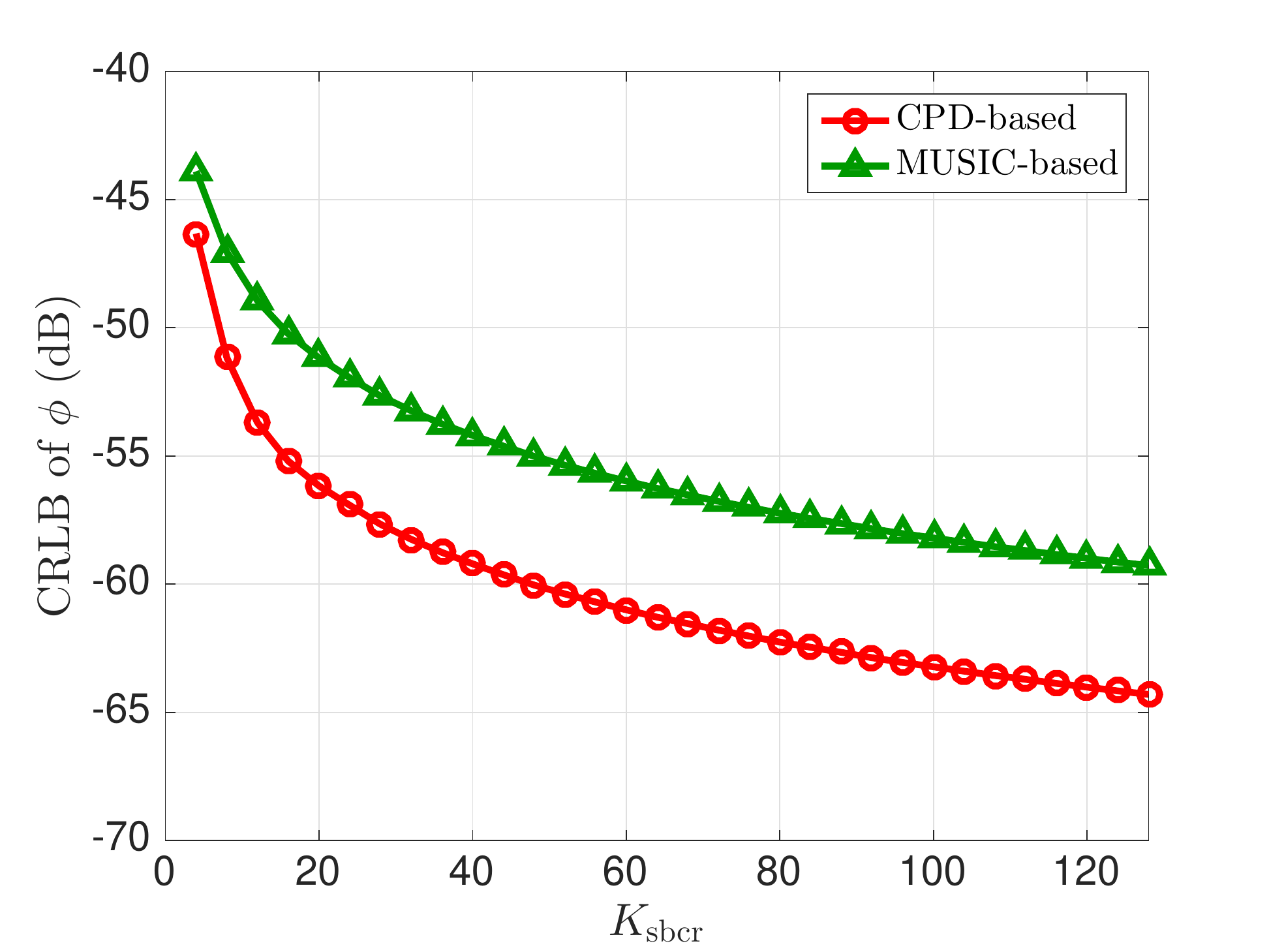}
		\label{fig:CRLB_phi_fixed_vsK}}
	\subfigure[center][{CRLB($\phi$) vs. $\Nrf$}]{
		\includegraphics[width=0.44\columnwidth]{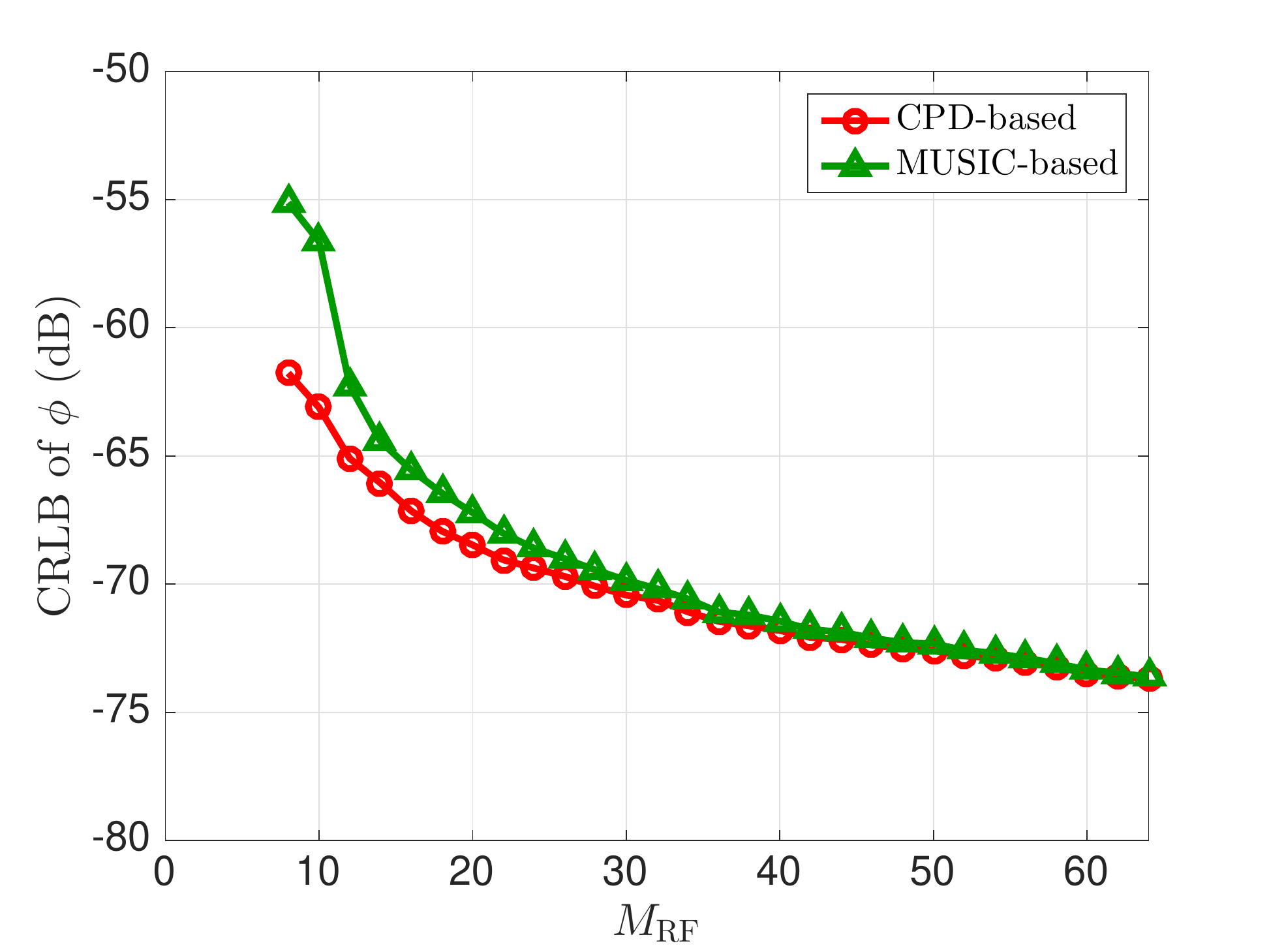}
		\label{fig:CRLB_phi_fixed_vsM}}
	\subfigure[center][{CRLB($\phi$) vs. SNR}]{
		\includegraphics[width=0.44\columnwidth]{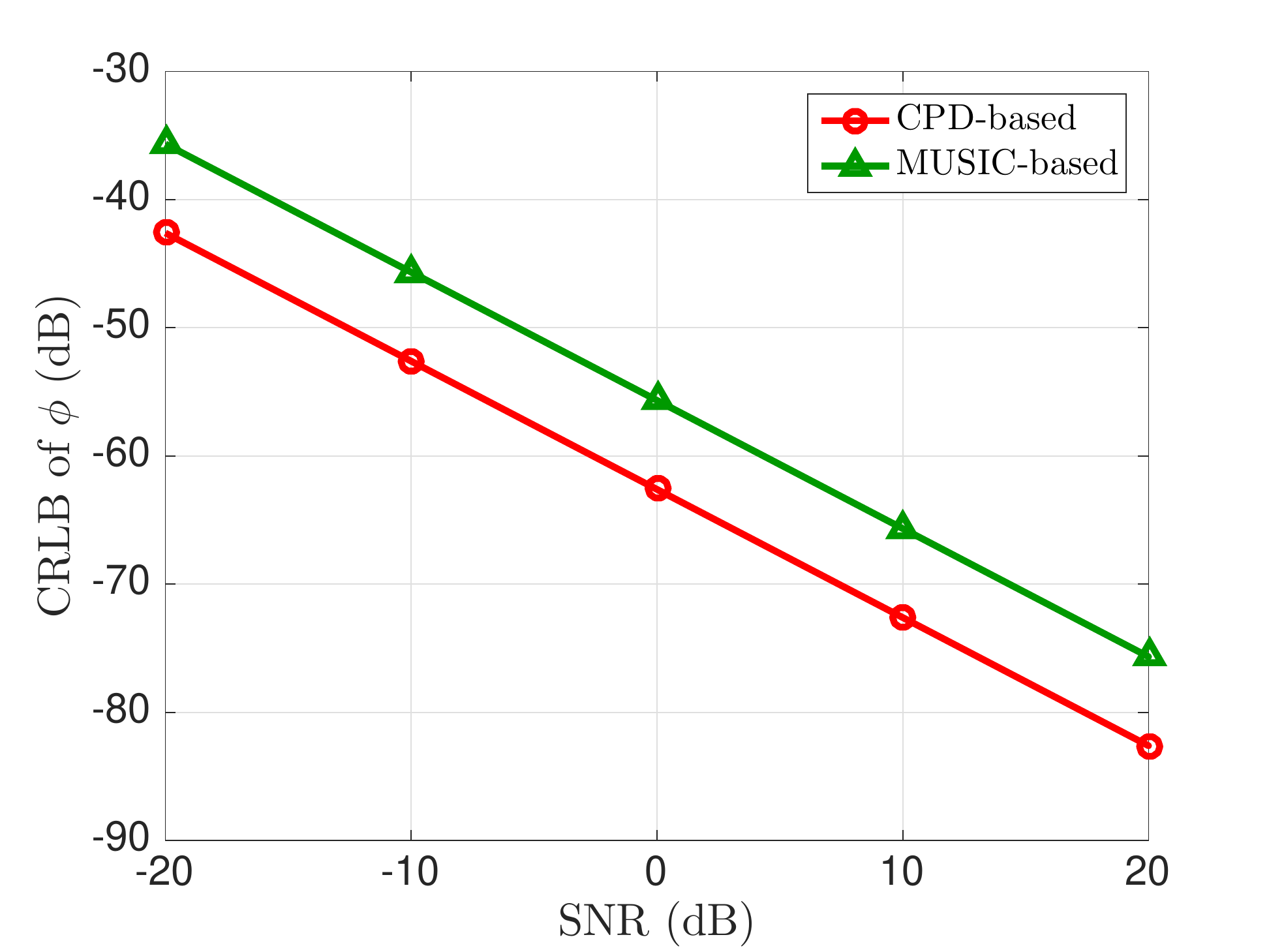}
		\label{fig:CRLB_phi_fixed_vsSNR}}
	\caption{ CRLB of $\phi$ with respect to deterministic $\boldsymbol{\phi}$, $\boldsymbol{\tau}$, and $\bG$ when $\Tt=20$, $ \Kf=128$, $\Nt=64$, $\Nrf=8$,  $\Lch=6$, and SNR 0 dB unless specified. }
	\label{fig:CRLBofPhi_vsT_vsK_vsM_vsSNR}
\end{figure}

In \figref{fig:comparison_MUSIC_performance_metric}, we show the MSE of the estimation of $\boldsymbol{\phi}$. We also compare the MSE results to the CRLB derived in \sref{sec:CRLB} for  $\Nt=64$, $\Nrf=8$, $\Lch=6$, $\Tt=20$, and $\Kf=128$. We assume that the path gains $g_{t,\ell}$'s are  generated from $\mathcal{CN}(0, 1/ \Lch)$ and $p_{\mathrm{PS}}(\tau) = \mathrm{sinc}(\tau / T_s)$. 
Since CRLB depends on the deterministic value of AoA and path delays, we set the values as $\boldsymbol{\phi} = [-66, 13, 49, -7, 81, 62]$ in degrees  and $\boldsymbol{\tau}/T_s = [0, 4.34, 7.13, 17.05, 21.08, 25.73] $ for the purpose of reproduction.
We can see that the proposed method achieves the MSE that is close to its theoretical lower bound at moderate and high SNR region.   


\begin{figure}[t]
	\centering
	\subfigure[center][{SNR 0 dB}]{
		\includegraphics[width=0.44\columnwidth]{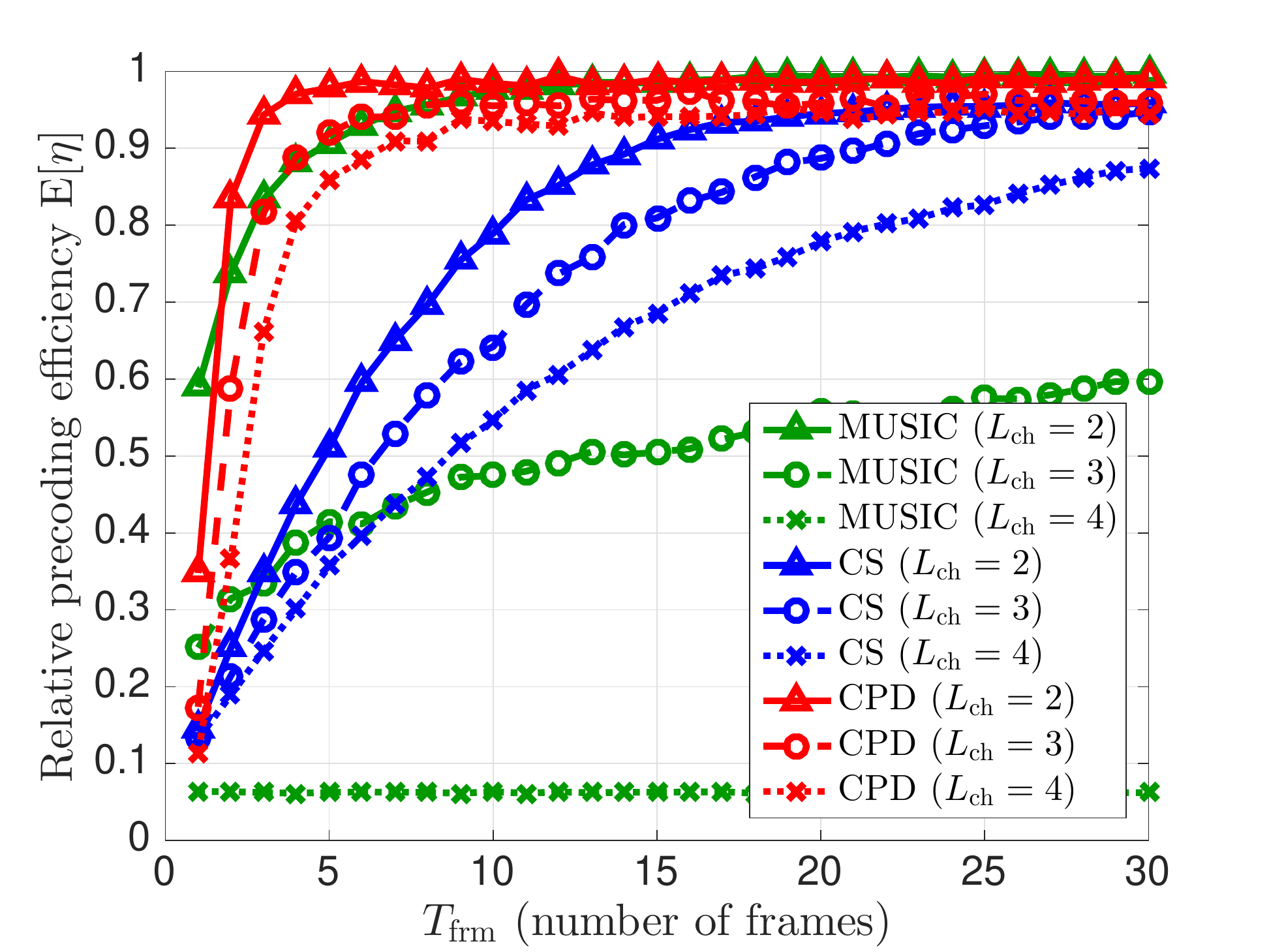}
		\label{fig:Eff_N64_M4_SNR0dB}}
	\subfigure[center][{SNR -10 dB}]{
		\includegraphics[width=0.44\columnwidth]{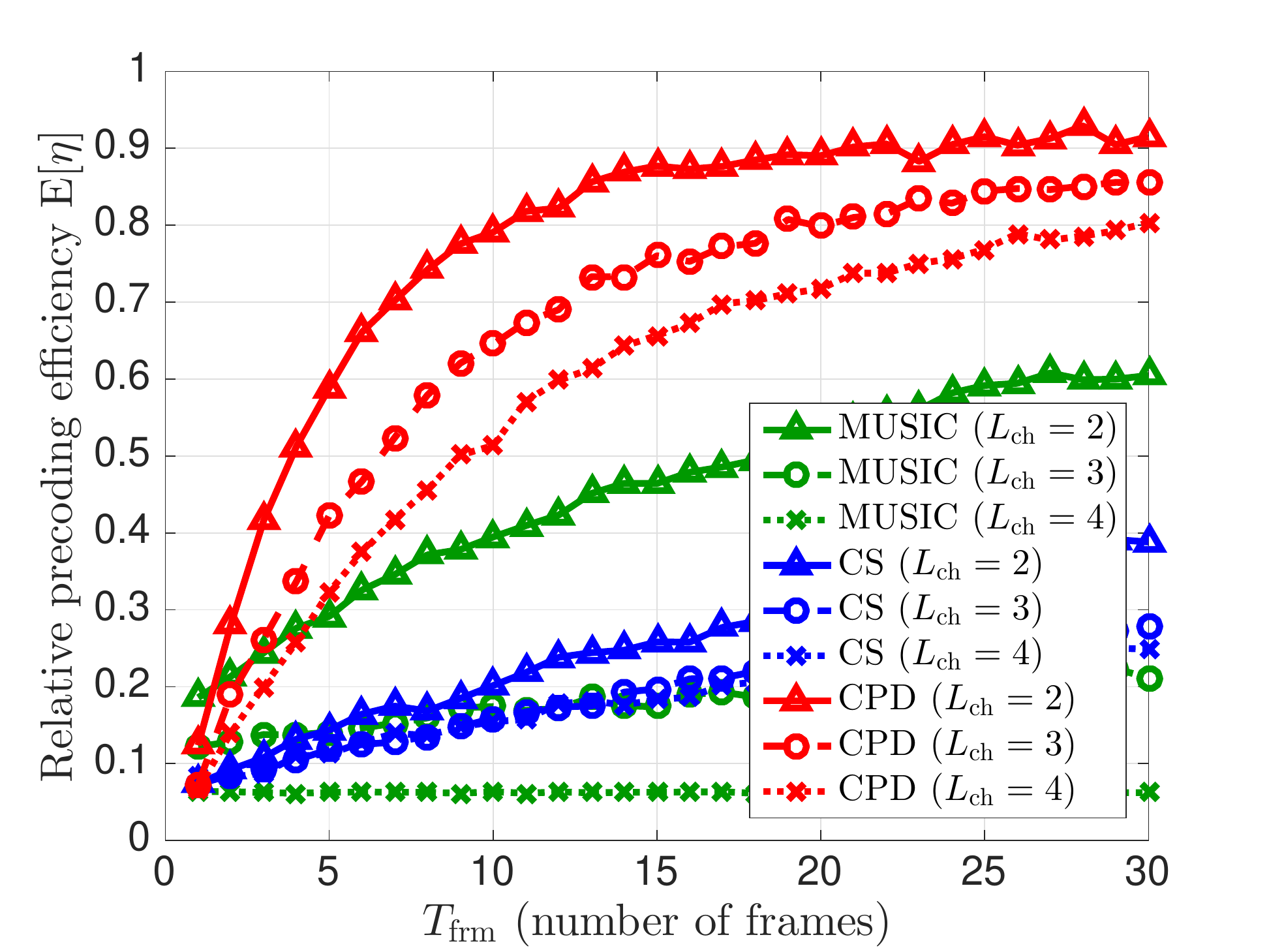}
		\label{fig:Eff_N64_M4_SNR-10dB}}
	\caption{ RPE vs. the number of frames $\Tt$ when $\Nt=64, \Nrf=4$, and $\Kf=128$.}
	\label{fig:Eff_N64_M4}
\end{figure}

\begin{figure}[t]
	\centering
	\subfigure[center][{SNR 0 dB}]{
		\includegraphics[width=0.44\columnwidth]{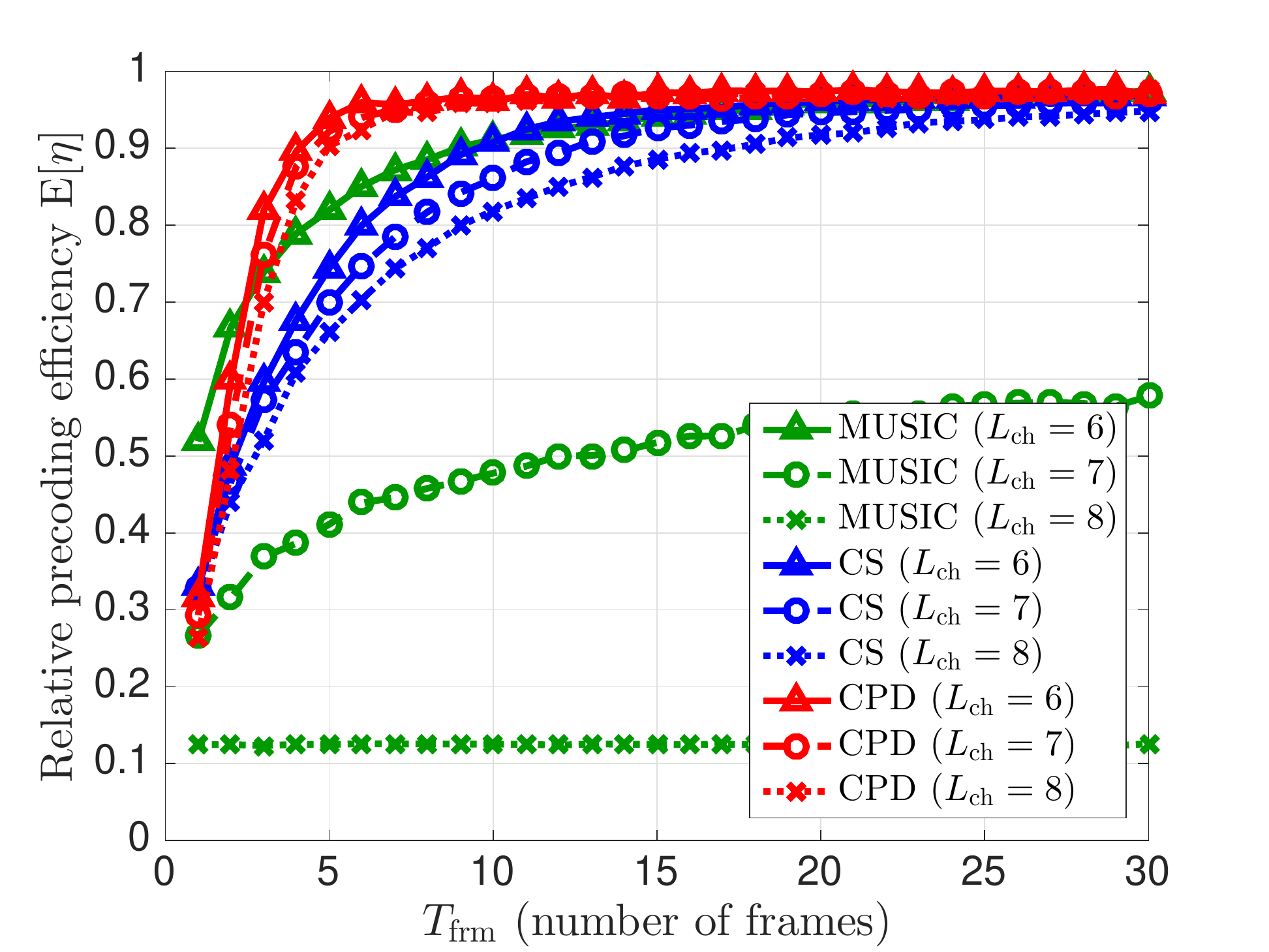}
		\label{fig:Eff_N64_M8_L678_SNR0dB}}
	\subfigure[center][{SNR -10 dB}]{
		\includegraphics[width=0.44\columnwidth]{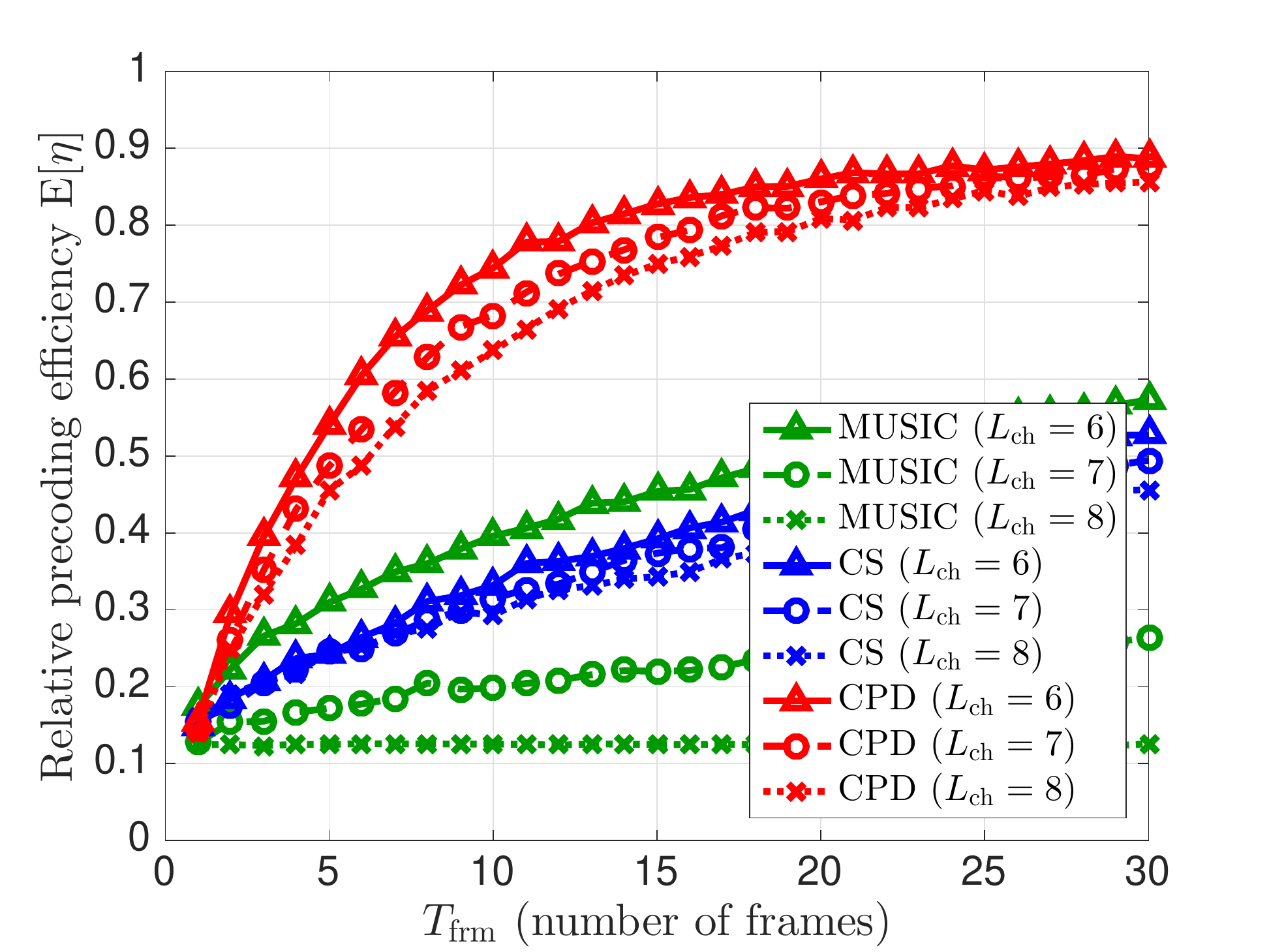}
		\label{fig:Eff_N64_M8_L678_SNR-10dB}}
	\caption{ RPE vs. the number of frames $\Tt$ when $\Nt=64, \Nrf=8$, and $\Kf=128$.}
	\label{fig:Eff_N64_M8}
\end{figure}

\figref{fig:CRLB_phi_wMUSIC} compares the proposed tensor-based method with  the MUSIC-based method in terms of the MSE($\phi$). In addition to numerical results, the analytical results indicate the superiority of the tensor-based method over the MUSIC-based method. The metric $1-\mathbb{E}[\eta]$ is plotted in \figref{fig:CRLB_EffMetric_wMUSIC} with the lower bound of its approximation derived in \sref{sec:Performance_metric}. As shown in \sref{sec:Performance_metric}, the RPE is closely related to MSE($\phi$) and its CRLB. 

\figref{fig:CRLBofPhi_vsT_vsK_vsM_vsSNR} shows the relationship between the CRLB of MSE($\phi$) and other system design parameters such as $\Tt$, $\Kf$, and $\Nrf$ as well as SNR. Unlike the relationship between  CRLB($\phi$) and SNR as shown in \figref{fig:CRLB_phi_fixed_vsSNR}, which is a linear relationship in dB, other parameters impact less on the CRLB as the parameter values increase. As shown in \figref{fig:CRLB_phi_fixed_vsM}, the CRLB of the proposed method approaches that of the MUSIC as $\Nrf$ increases.

\subsection{Performance evaluation of the spatial channel covariance estimation methods }\label{ssec::simulation_eff}

In this subsection, we evaluate the performance of the spatial channel covariance estimation in terms of the RPE  for  $\Nt=64$ and $\Kf=128$. Unlike \sref{ssec::CRLB_sim}, the AoA $\phi_{\ell}$s are   uniformly distributed in $[-180^{\circ}, 180^{\circ}]$,  and the normalized delay $\tau_{\ell}/T_s$s are uniformly distributed in  $[0, N_{\mathrm{CP}}]$, where the cyclic prefix length $N_{\mathrm{CP}}$ is set to $\Kf/4$.
The path gains $g_{t,\ell}$'s are IID complex Gaussian random variables as $g_{t,\ell} \sim \mathcal{CN}(0, 1/ \Lch)$ and $p_{\mathrm{PS}}(\tau) = \mathrm{sinc}(\tau / T_s)$.

\figref{fig:Eff_N64_M4} compares the proposed method with the two other methods explained in \sref{sec:Comparison_others} in terms of the RPE when $\Nrf=4$. The two figures,  \figref{fig:Eff_N64_M4_SNR0dB}, and \figref{fig:Eff_N64_M4_SNR-10dB}, show the comparison at different SNR values: 0 dB and -10 dB. In each figure, we compare the methods for different $\Lch$ values: 2, 3, and 4. 
It is worthwhile to note that the MUSIC-based method does not work properly if $\Lch \geq \Nrf$  because $\bU_{\bn}$  in \eqref{MUSIC_function} must have at least one column. 
Even when $\Lch < \Nrf$, the figure shows that the performance degradation of the MUSIC-based method is more severe than other two methods as $\Lch$ approaches $\Nrf$. In addition, as SNR becomes low, the RPE of the MUSIC-based method rapidly decreases compared to that of the tensor-based method. 
The CS-based method outperforms the MUSIC-based method when $\Lch$ relative to $\Nrf$ is large, but the method shows poor performance at very low SNR. The proposed method based on the higher-tensor CPD shows the best performance in most cases, in particular when $\Lch$ is not so small and SNR is low. \figref{fig:Eff_N64_M8} shows the results when $\Nrf=8$, and the results have the same trend as in   \figref{fig:Eff_N64_M4} where $\Nrf=4$.

\begin{figure}[t]
	\centering
	\subfigure[center][{SNR 0 dB}]{
		\includegraphics[width=0.44\columnwidth]{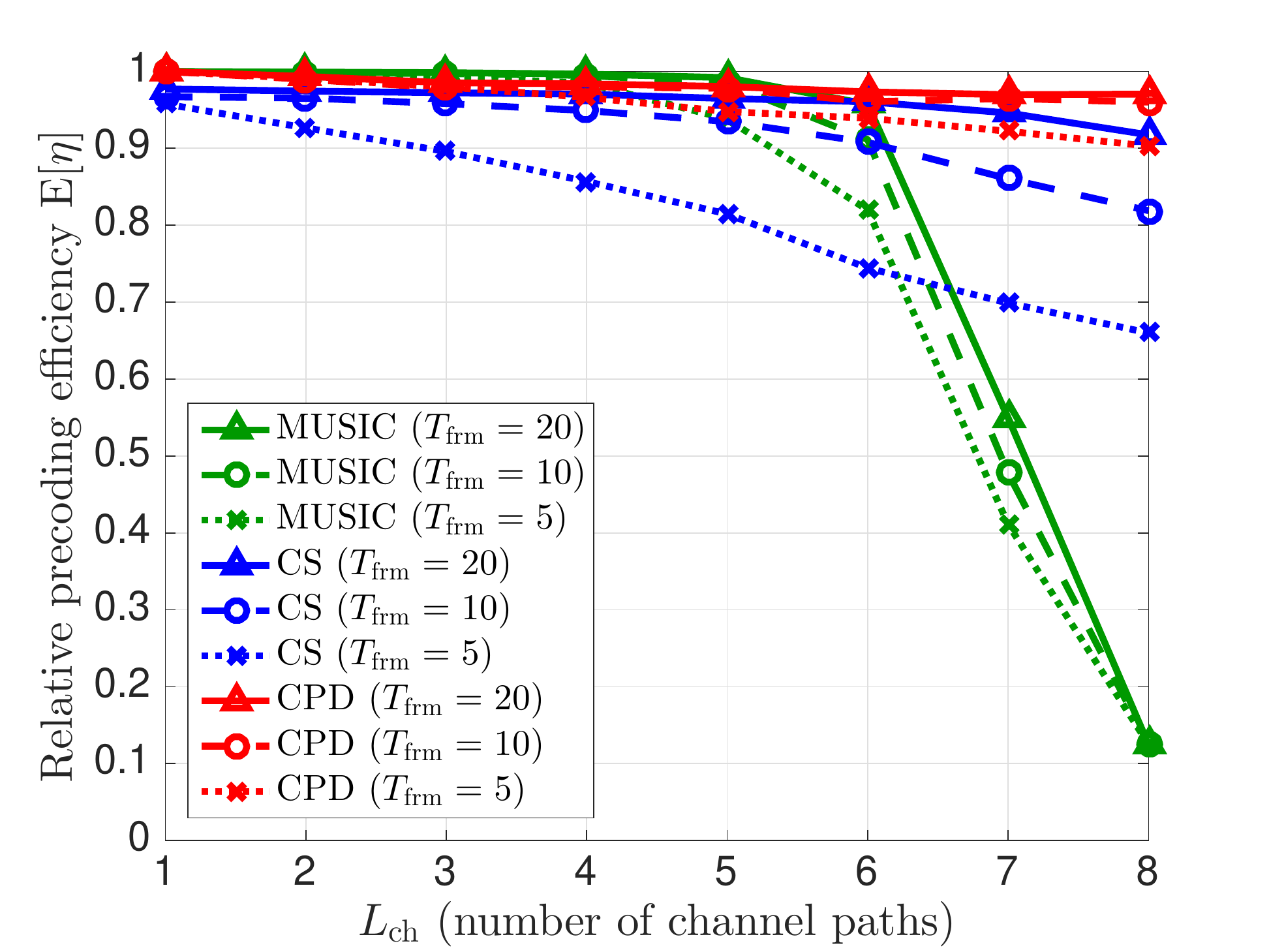}
		\label{fig:Eff_vsL_N64_M8_SNR0dB}}
	\subfigure[center][{$\Tt=20$}]{
		\includegraphics[width=0.44\columnwidth]{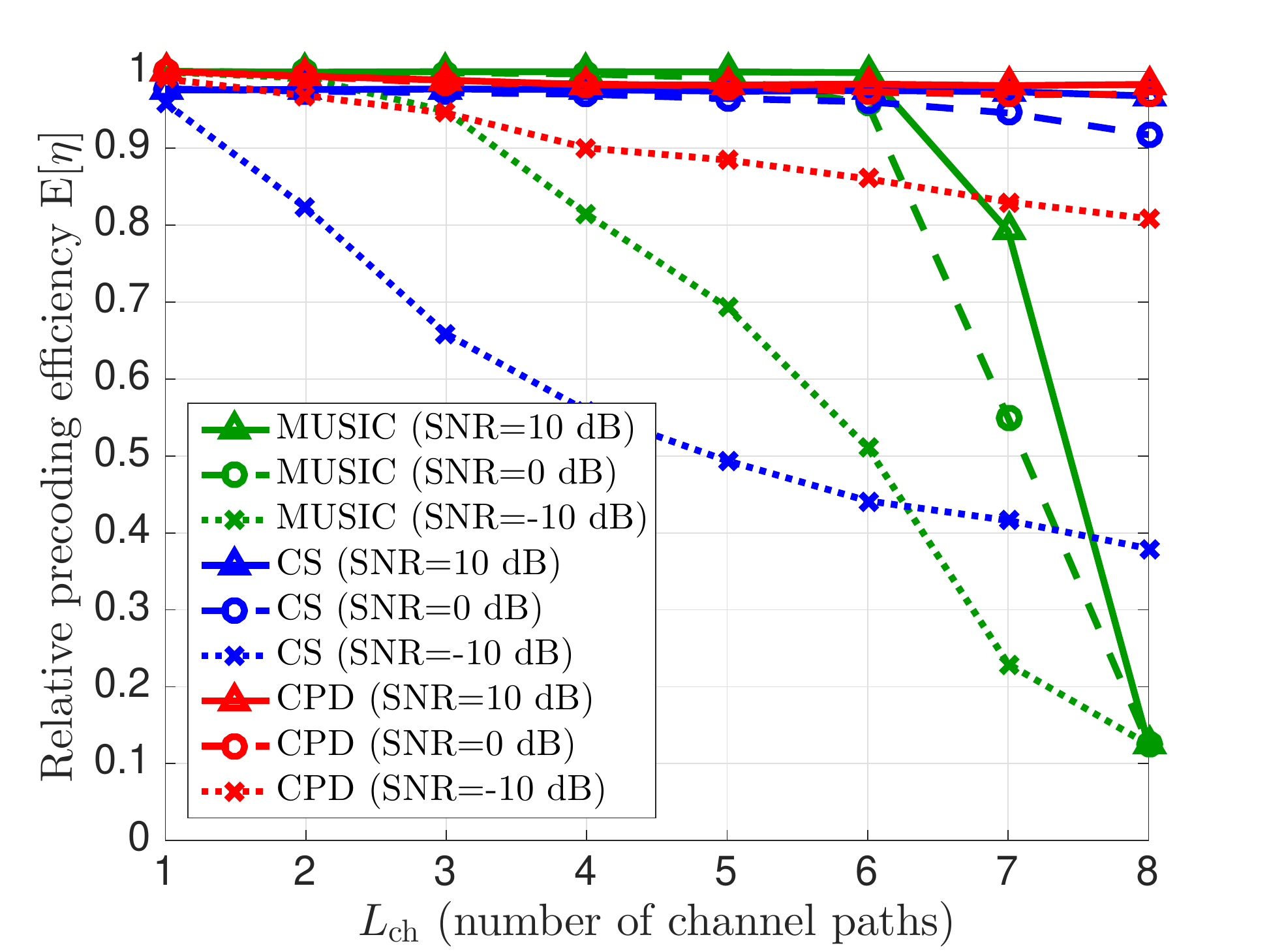}
		\label{fig:Eff_vsL_N64_M8_T20}}
	\caption{ RPE vs. the number of channel paths $\Lch$ when $\Nt=64, \Nrf=8$, and $\Kf=128$.}
	\label{fig:Eff_vsL_N64_M8}
\end{figure}

\figref{fig:Eff_vsL_N64_M8} shows the dependency of the RPE on the number of channel paths $\Lch$  for each method. \figref{fig:Eff_vsL_N64_M8_SNR0dB} reveals that the performance of the MUSIC-based method rapidly decreases after $\Lch$ becomes larger than six. In \figref{fig:Eff_vsL_N64_M8_T20}, the number of frames $\Tt$ is fixed at 20, and the RPEs are compared for different SNR values. For SNR 0 dB and 10 dB, the CS-based method has as a high RPE as the tensor-based method, but its performance is significantly degraded as SNR goes to  -10 dB. For any $\Lch$ values, we can observe that the proposed tensor-based method has a  reasonably high RPE. 


\begin{figure}[t]
	\centering
	\subfigure[center][{SNR 0 dB}]{
		\includegraphics[width=0.44\columnwidth]{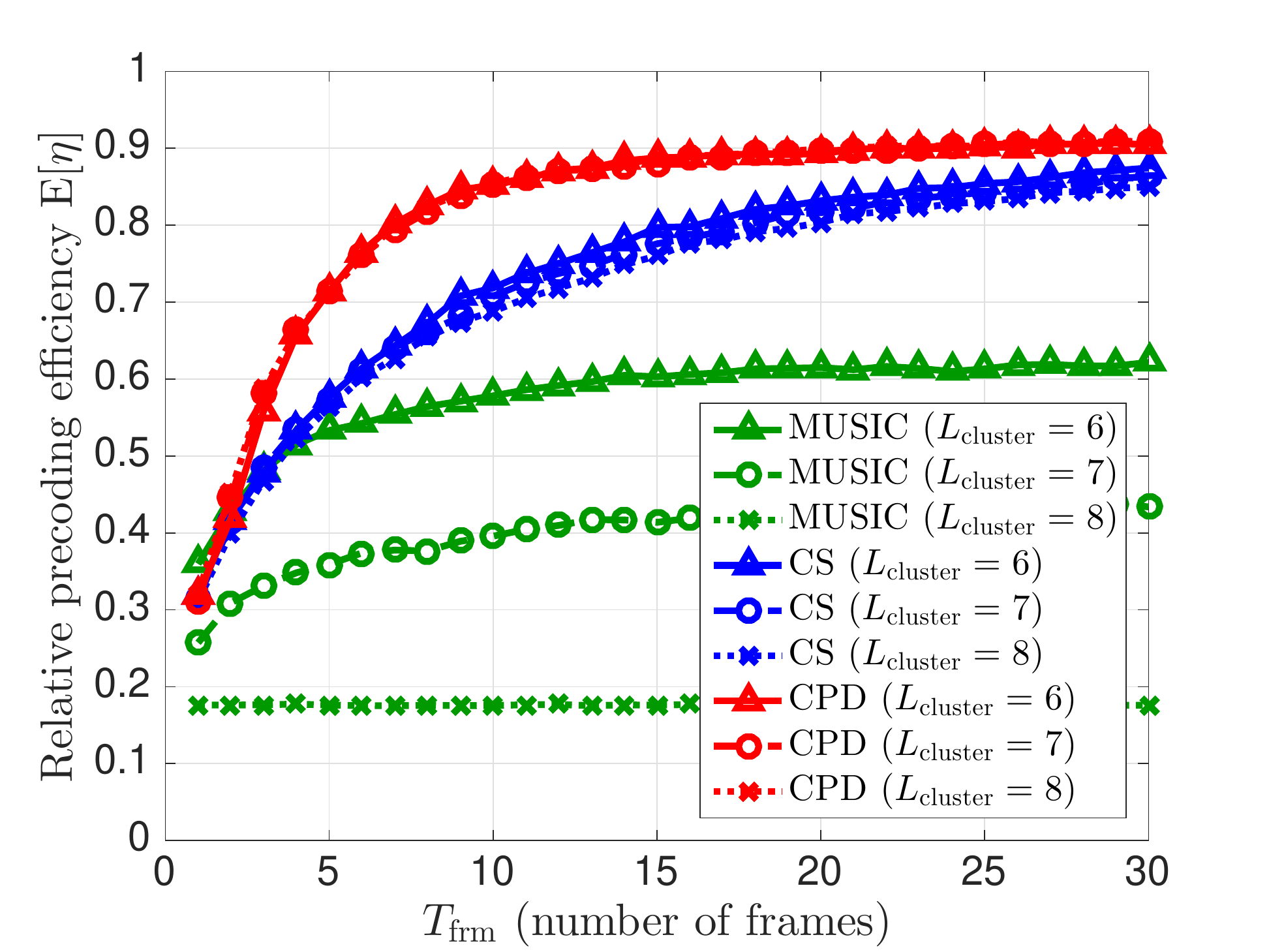}
		\label{fig_eff_metric_N64_M8_Lcluster678_Lrays10_SNR0dB}}
	\subfigure[center][{SNR -10 dB}]{
		\includegraphics[width=0.44\columnwidth]{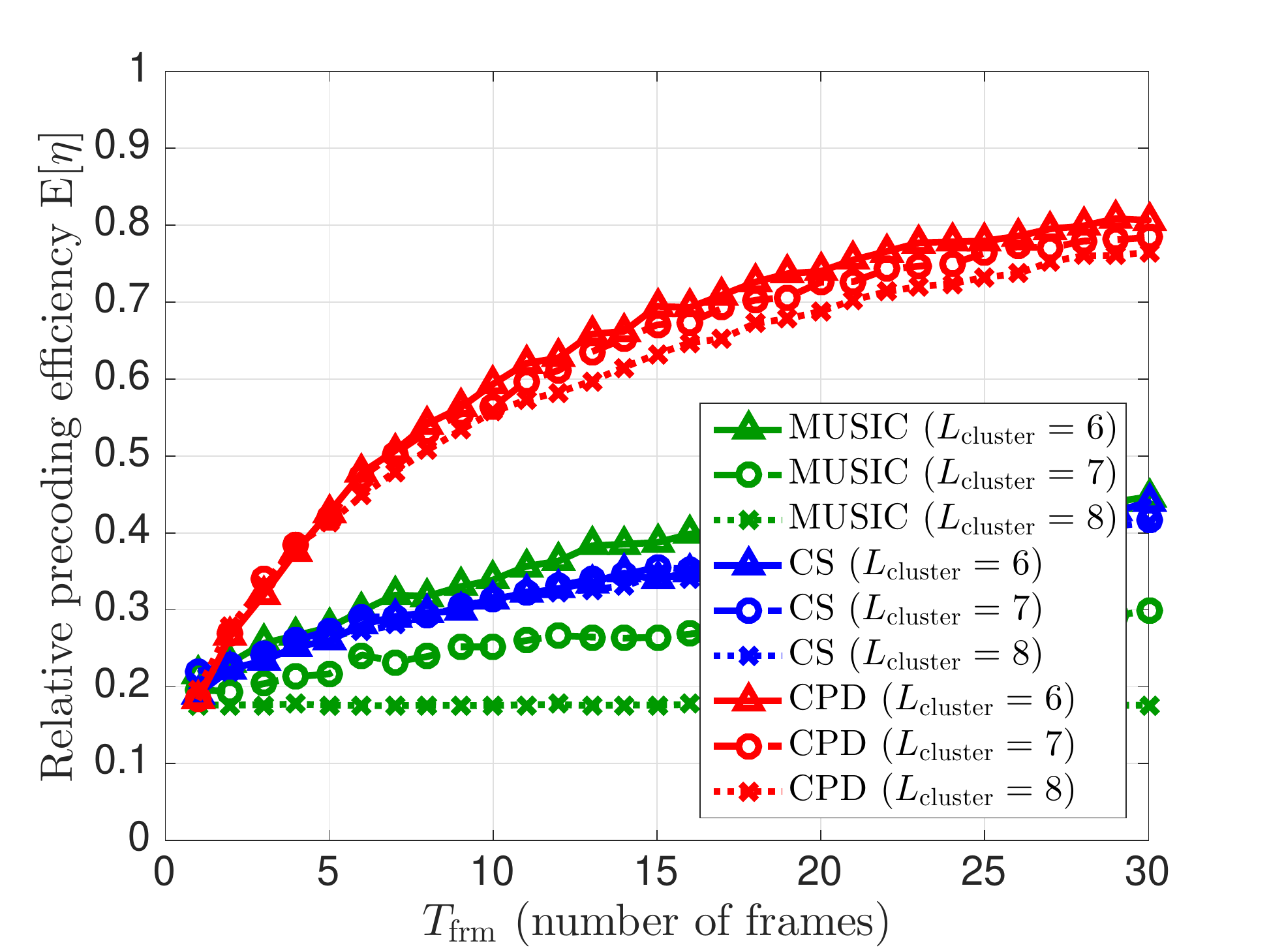}
		\label{fig_eff_metric_N64_M8_Lcluster678_Lrays10_SNRm10dB}}
	\caption{ RPE vs. the number of frames $\Tt$ when $\Nt=64, \Nrf=8$, and $\Kf=128$. The channel has $L_{\mathrm{cluster}}=6,7,$ or $8$ clusters, and each cluster has $L_{\mathrm{subray}}=10$ subrays with angular spread $2^{\circ}$.}
	\label{fig:Eff_vsT_multiple_subrays}
\end{figure}

Until now, we assumed that the channel has only $\Lch$ channel paths as in \eqref{WB_ch_model_CIR}. 
To evaluate performance for more realistic channels, we consider a clustered channel model \cite{WINNER2} that has multiple clusters with multiple subrays as 
\begin{equation}\label{WB_ch_model_CIR_cluster}
\bh_{t}[d]  = \sum_{\ell_c=1}^{L_{\mathrm{cluster}}} \sum_{\ell_s=1}^{L_{\mathrm{subray}}}  g_{t,\ell_c,\ell_s} p_{\mathrm{PS}}(dT_s - \tau_{\ell_c,\ell_s}) \ba ( \phi_{\ell_c,\ell_s}).
\end{equation}
The channel has $L_{\mathrm{cluster}}$ clusters whose AoAs are uniformly distributed in $[-180^{\circ}, 180^{\circ} ]$, and each cluster has $L_{\mathrm{subray}}$ subrays whose AoA offsets are Laplacian distributed with angular spread $2^{\circ}$. 
All subrays within a cluster are assumed to have the same delay.  
Although the rank of the channel tensor $ L_{\mathrm{cluster}} L_{\mathrm{subrays}}$ becomes  high in general, we can approximate the channel tensor to be a low-rank tensor for spatially sparse channels. 
\figref{fig:Eff_vsT_multiple_subrays} shows the RPE results when $\Nt = 64$, $\Nrf = 8$, $\Kf = 128$,  $ L_{\mathrm{subray}}=10$, and $ L_{\mathrm{cluster}}=6,7$, or $8$. 
Instead of using the actual number of channel paths $\Lch = L_{\mathrm{cluster}} L_{\mathrm{subray}}$,  the methods based on CPD and CS use $\Nrf$ for its low-rank (or sparse) approximation  while the MUSIC-based method uses $L_{\mathrm{cluster}}$ due to its inherent limitation of using $\Nrf$.  
The figure shows that, although the multiple subrays result in performance loss compared to the single subray case, the proposed method still works properly even in this case. 

\section{Conclusions}\label{sec:conclusions}

In this paper, we proposed a spatial channel covariance estimation method for the hybrid analog/digital architecture over time-varying frequency-selective channels. Leveraging the fact that a low-rank higher-order tensor can be uniquely decomposed into factor matrices in each domain, we formulated the estimation problem by using high-order tensors and proposed a solution that achieves performance close to its theoretical bound. 
We also derived the CRLB of the proposed method and showed that compared it is lower than the CRLB of MUSIC-based approach.
Numerical results showed that our proposed work outperforms the MUSIC-based work and the CS-based work. 
The results also showed that our proposed work has a more significant gain in the low SNR regime and the performance degradation caused by the increase in the number of channel paths is less severe than prior work.

\bibliographystyle{IEEEtran}
\bibliography{IEEEabrv,Bib_cov_est_tensor_r1}

\end{document}